\documentstyle[12pt,epsf,epsfig]{article}
\newcount\timecount
\newcount\hours \newcount\minutes  \newcount\temp \newcount\pmhours
\hours = \time
\divide\hours by 60
\temp = \hours
\multiply\temp by 60
\minutes = \time
\advance\minutes by -\temp
\def\hour{\the\hours}
\def\minute{\ifnum\minutes<10 0\the\minutes
            \else\the\minutes\fi}
\def\clock{
\ifnum\hours=0 12:\minute\ AM
\else\ifnum\hours<12 \hour:\minute\ AM
      \else\ifnum\hours=12 12:\minute\ PM
            \else\ifnum\hours>12
                 \pmhours=\hours
                 \advance\pmhours by -12
                 \the\pmhours:\minute\ PM
                 \fi
            \fi
      \fi
\fi
}

\def\monthname{\relax\ifcase\month 0/\or January\or February\or
   March\or April\or May\or June\or July\or August\or September\or
   October\or November\or December\else\number\month/\fi}

\def\bold#1{\setbox0=\hbox{$#1$}%
     \kern-.025em\copy0\kern-\wd0
     \kern.05em\copy0\kern-\wd0
     \kern-.025em\raise.0433em\box0 }

\def\beq{\begin{equation}}
\def\eeq{\end{equation}}

\def\ga{\mathrel{\raise.3ex\hbox{$>$\kern-.75em\lower1ex\hbox{$\sim$}}}}
\def\la{\mathrel{\raise.3ex\hbox{$<$\kern-.75em\lower1ex\hbox{$\sim$}}}}
\def\gev{{\rm \, Ge\kern-0.125em V}}
\def\tev{{\rm \, Te\kern-0.125em V}}
\def\gyr{{\rm \, G\kern-0.125em yr}}

\def\gappeq{\mathrel{\rlap {\raise.5ex\hbox{$>$}}
{\lower.5ex\hbox{$\sim$}}}}
\def\lappeq{\mathrel{\rlap{\raise.5ex\hbox{$<$}}
{\lower.5ex\hbox{$\sim$}}}}
\def\Toprel#1\over#2{\mathrel{\mathop{#2}\limits^{#1}}}

\def\m12{m_{1\!/2}}

\begin{document}
\begin{titlepage}
\pagestyle{empty}
\baselineskip=21pt
\rightline{\tt hep-ph/0310356}
\rightline{CERN--TH/2003-262}
\rightline{UMN--TH--2217/03}
\rightline{TPI--MINN--03/28}
\vskip 0.2in
\begin{center}
{\large {\bf Likelihood Analysis of the CMSSM Parameter Space}}
\end{center}
\begin{center}
\vskip 0.2in
{\bf John~Ellis}$^1$, {\bf Keith~A.~Olive}$^{2}$, {\bf Yudi Santoso}$^{2}$ 
and {\bf Vassilis~C. Spanos}$^{2}$
\vskip 0.1in
{\it
$^1${TH Division, CERN, Geneva, Switzerland}\\
$^2${William I. Fine Theoretical Physics Institute, \\
University of Minnesota, Minneapolis, MN 55455, USA}}\\
\vskip 0.2in
{\bf Abstract}
\end{center}
\baselineskip=18pt \noindent

We present a likelihood analysis of the parameter space of the constrained
minimal supersymmetric extension of the Standard Model (CMSSM), in which
the input scalar masses $m_0$ and fermion masses $m_{1/2}$ are each
assumed to be universal. We include the full experimental likelihood
function from the LEP Higgs search as well as the likelihood from a global
precision electroweak fit.  We also include the likelihoods for $b \to s
\gamma$ decay and (optionally) $g_\mu - 2$. For each of these inputs, both
the experimental and theoretical errors are treated. We include the
systematic errors stemming from the uncertainties in $m_t$ and $m_b$,
which are important for delineating the allowed CMSSM parameter space as
well as calculating the relic density of supersymmetric particles. We
assume that these dominate the cold dark matter density, with a density in
the range favoured by WMAP. We display the global likelihood function
along cuts in the $(m_{1/2}, m_0)$ planes for $\tan \beta = 10$ and both
signs of $\mu$, $\tan \beta = 35, \mu < 0$ and $\tan \beta = 50, \mu > 0$,
which illustrate the relevance of $g_\mu - 2$ and the uncertainty in
$m_t$. We also display likelihood contours in the $(m_{1/2}, m_0)$ planes
for these values of $\tan \beta$. The likelihood function is generally
larger for $\mu > 0$ than for $\mu < 0$, and smaller in the focus-point
region than in the bulk and coannihilation regions, but none of these
possibilities can yet be excluded.

\vfill
\leftline{CERN--TH/2003-262}
\leftline{October 2003}
\end{titlepage}
\baselineskip=18pt

\section{Introduction}

Supersymmetry remains one of the best-motivated frameworks for possible
physics beyond the Standard Model, and many analyses have been published
of the parameter space of the minimal supersymmetric extension of the
Standard Model (MSSM). It is often assumed that  the soft
supersymmetry-breaking mass terms $m_{1/2}, m_0$ are universal at an input
GUT scale, a restriction referred to as the constrained MSSM (CMSSM). In 
addition to
experimental constraints from sparticle and Higgs searches at 
LEP~\cite{LEPHWG}, the
measured rate for $b \to s \gamma$~\cite{bsgex} and the value of $g_\mu -
2$~\cite{BNL}~\footnote{In view of the chequered history of this 
constraint, we
present results obtained neglecting $g_\mu - 2$, as well as results using
the latest re-evaluation of the Standard Model
contribution~\cite{newDavier}.}, the CMSSM parameter space is also
restricted by the cosmological density of non-baryonic cold dark matter,
$\Omega_{CDM}$ \cite{ganis,LS,eos2,cmssm}. It is also often assumed that most
of $\Omega_{CDM}$ is
provided by the lightest supersymmetric particle (LSP), which we presume
to be the lightest neutralino $\chi$.

The importance of cold dark matter has recently been supported by the WMAP
Collaboration~\cite{wmap,reion}, which has established a strong upper limit on
hot dark matter in the form of neutrinos. Moreover, the WMAP Collaboration
also reports the observation of early reionization when $z \sim
20$~\cite{reion}, which disfavours models with warm dark matter.
Furthermore, the WMAP data greatly restrict the possible range for the
density of cold dark matter: $\Omega_{CDM} h^2 =
0.1126^{+0.0081}_{-0.0091}$ (one-$\sigma$ errors). Several recent papers
have combined this information with experimental constraints on the CMSSM
parameter space~\cite{EOSS,LN,Baer,Nath,Arnowitt}, assuming that LSPs 
dominate $\Omega_{CDM}$.

The optimal way to combine these various constraints is via a likelihood
analysis, as has been done by some authors both before~\cite{DeBoer} and
after~\cite{Baer} the WMAP data was released. When performing such an analysis,
in addition to the formal experimental errors, it is also essential to
take into account theoretical errors, which introduce systematic
uncertainties that are frequently non-negligible. The main aim of this
paper is to present a new likelihood analysis which includes a careful
treatment of these errors.

The precision of the WMAP constraint on $\Omega_{CDM}$ selects narrow
strips in the CMSSM parameter space, even in the former `bulk' region at
low $m_{1/2}$ and $m_0$. This narrowing is even more apparent in the
coannihilation `tail' of parameter space extending to larger $m_{1/2}$, in
the `funnels' due to rapid annihilations through the $A$ and $H$ poles
that appear at large $\tan \beta$, and in the focus-point region at large
$m_0$, close to the boundary of the area where electroweak symmetry
breaking remains possible. The experimental and theoretical errors are
crucial for estimating the widths of these narrow strips, and also for
calculating the likelihood function along cuts across them, as well as for
the global likelihood contours we present in the $(m_{1/2}, m_0)$ planes
for different choices of $\tan \beta$ and the sign of $\mu$.

In the `bulk' and coannihilation regions, we find that the theoretical
uncertainties are relatively small, though they could become dominant if
the experimental error in $\Omega_{CDM} h^2$ is reduced below 5\% some
time in the future. However, theoretical uncertainties in the calculation
of $m_h$ do have an effect on the lower end of the `bulk' region, and
these are sensitive to the experimental and theoretical uncertainties in
$m_t$ and (at large $\tan \beta$) also $m_b$. The theoretical errors due
to the current uncertainties in $m_b$ and $m_t$ are dominant in the
`funnel' and `focus-point' regions, respectively. These sensitivities may
explain some of the discrepancies between the results of different codes
for calculating the supersymmetric relic density, which are particularly
apparent in these regions. These sensitivities imply that results depend
on the treatment of higher-order effects, for which there are not always
unique prescriptions.

With our treatment of these uncertainties, we find that the half-plane
with $\mu > 0$ is generally favoured over that with $\mu < 0$, and that,
within each half-plane, the coannihilation region of the CMSSM parameter
space is generally favoured over the focus-point region~\footnote{Our
conclusions differ in this respect from those of~\cite{Baer}.}, but these
preferences are not strong.

The rest of the paper is organized as follows. In section~\ref{sec:treat},
we discuss the treatment of the various constraints employed to
define the global likelihood function. In section~\ref{sec:strips}, we
present the profile of the global likelihood function along cuts in the
$(m_{1/2},m_0)$ plane for different choices of $\tan\beta$ and the sign of
$\mu$.  In section~\ref{sec:contours}, we present iso-likelihood contours
at certain CLs, obtained by integrating the likelihood function. Finally,
in section~\ref{sec:summary}, we summarize our findings and suggest
directions for future analyses of this type.

\section{Constraints on the CMSSM Parameter Space}
\label{sec:treat}

\subsection{Particle Searches}

We first discuss the implementation of the accelerator constraints on
CMSSM particle masses. Previous studies have shown that the LEP limits on
the masses of sparticles such as the selectron and chargino constrain the
CMSSM parameter space much less than the LEP Higgs limit and $b \to s
\gamma$ (see, e.g.,~\cite{eos2,benchmark}).
As we have discussed previously, in the CMSSM parameter regions
of interest, the LEP Higgs constraint reduces essentially to that on the
Standard Model Higgs boson \cite{benchmark}. This is often implemented as the 95\%
confidence-level lower limit $m_h > 114.4$~GeV~\cite{LEPHWG}. However,
here we use the full likelihood function for the LEP Higgs search, as
released by the LEP Higgs Working Group. This includes the small
enhancement in the likelihood just beyond the formal limit due to the LEP
Higgs signal reported late in 2000. This was re-evaluated most recently
in~\cite{LEPHWG}, and cannot be regarded as significant evidence for a
light Higgs boson. We have also taken into account the indirect information 
on $m_h$ provided by a global fit to the precision electroweak data. 
The likelihood function from this indirect source does not vary 
rapidly over the range of Higgs masses found in the CMSSM, but we 
include this contribution with the aim of completeness.

The interpretation of the combined Higgs likelihood, ${\cal L}_{exp}$,  
in the $(m_{1/2},
m_0)$ plane depends on uncertainties in the theoretical calculation of
$m_h$. These include the experimental error in $m_t$ and (particularly at
large $\tan \beta$) $m_b$, and theoretical uncertainties associated with
higher-order corrections to $m_h$. Our default assumptions are that $m_t =
175 \pm 5$~GeV for the pole mass, and $m_b = 4.25 \pm 0.25$~GeV for the
running $\overline {MS}$ mass evaluated at $m_b$ itself.
The theoretical uncertainty in $m_h$,  $\sigma_{th}$,  is dominated by
the experimental uncertainties in $m_{t,b}$, which are 
treated as uncorrelated Gaussian errors:
\beq
\sigma_{th}^2 = \left( \frac{\partial m_h}{\partial m_t} \right)^2 \Delta 
m_t^2 + \left( \frac{\partial m_h}{\partial m_b} \right)^2 \Delta m_b^2 \,.
\label{eq:sigmath}
\eeq
The Higgs mass is calculated using  the  latest version
of {\tt FeynHiggs}~\cite{FeynHiggs}. 
Typically, we find that $(\partial m_h/\partial m_t) \sim 0.5$, so that 
$\sigma_{th}$ is roughly 2-3 GeV.
Subdominant two-loop contributions as well as  higher-order 
corrections have been shown to contribute much less~\cite{martin}.

The
combined experimental likelihood, ${\cal L}_{exp}$, from 
direct searches at LEP~2 and a global
electroweak fit is then convolved with a theoretical likelihood
(taken as a Gaussian) with uncertainty given by $\sigma_{th}$ from
(\ref{eq:sigmath}) above. Thus, we define the
total Higgs likelihood function, ${\cal L}_h$, as
\beq
{\cal L}_h(m_h) = { {\cal N} \over {\sqrt{2 \pi}\, \sigma_{th}  }}
 \int d m^{\prime}_h \,\, {\cal L}_{exp}(m^{\prime}_h)
 \,\, e^{-(m^{\prime}_h-m_h)^2/2 \sigma_{th}^2 }\, ,
\label{eq:higlik}
\eeq
where ${\cal N}$ is a factor that normalizes  the experimental likelihood
distribution. 

\subsection{$b \to s \gamma$ Decay}

The branching ratio for the rare decays $b \to s \gamma$ has been measured
by the CLEO, BELLE and BaBar collaborations~\cite{bsgex}, and we take as
the combined value ${\cal{B}}(b \to s \gamma)=(3.54 \pm 0.41 \pm
0.26)\times 10^{-4}$. The theoretical prediction of $b \to s \gamma$
\cite{gam,bsgth} contains uncertainties which stem from the uncertainties
in $m_b$, $\alpha_s$, the measurement of the semileptonic branching ratio
of the $B$ meson as well as the effect of the scale dependence. In
particular, the scale dependence of the theoretical prediction arises from
the dependence on three scales: the scale where the QCD corrections to the
semileptonic decay are calculated and the high and low energy scales,
relevant to $b \to s \gamma$ decay. These sources of uncertainty can be
combined to determine a total theoretical uncertainty. Finally, the
experimental measurement is converted into a Gaussian likelihood and
convolved with a theoretical likelihood to determine the total likelihood
${\cal L}_{bsg}$ containing both experimental and theoretical
uncertainties~\cite{gam}~\footnote{Further details of our treatment of
experimental and theoretical errors, as applied to the CMSSM, can be found
in~\cite{ganis}.}.

\subsection{Measurement of $g_\mu - 2$}

The interpretation of the BNL measurement of $a_\mu \equiv g_\mu - 
2$~\cite{BNL} is 
not yet settled. Two updated Standard Model predictions for $a_\mu$ 
have recently been calculated \cite{newDavier}. One is based on $e^+ e^- \to$~hadrons data, 
incorporating the recent re-evaluation of radiative cross 
sections by the CMD-2 group: 
\begin{equation}
a_\mu = (11,659,180.9 \pm 7.2 \pm 3.5 \pm 0.4) \times 10^{-10},
\label{amuhad}
\end{equation}
and the second estimate is based on $\tau$ decay data:
\begin{equation}
a_\mu = (11,659,195.6 \pm 5.8 \pm 3.5 \pm 0.4) \times 10^{-10},
\label{amutau}
\end{equation}
where, in each case, the first error is due to uncertainties in the
hadronic vacuum polarization, the second is due to light-by-light
scattering and the third combines higher-order QED and electroweak
uncertainties. Comparing these estimates with the experimental 
value~\cite{BNL}, one 
finds discrepancies
\begin{equation}
\Delta a_\mu = (22.1 \pm 7.2 \pm 3.5 \pm 8.0) \times 10^{-10} (1.9~\sigma)
\label{delta}
\end{equation}
and
\begin{equation}
\Delta a_\mu = (7.4 \pm 5.8 \pm 3.5 \pm 8.0) \times 10^{-10} (0.7~\sigma),
\label{deltatau}
\end{equation}
for the $e^+ e^-$ and $\tau$ estimates, respectively, where the 
second error is from the light-by-light scattering contribution and the 
last is the experimental error from the BNL measurement.

Based on the $e^+ e^-$ estimate, one would tempted to think there is some 
hint for new physics beyond the Standard Model. However, the $\tau$ 
estimate does not confirm this optimistic picture. Awaiting 
clarification of the discrepancy between the $e^+ e^-$ and $\tau$ data, 
we calculate the likelihood function for the CMSSM under two 
hypotheses:
\begin{itemize}
\item{neglecting any information from $g_\mu - 2$, which may be 
unduly pessimistic, and}
\item{taking the $e^+ e^-$ estimate (\ref{delta}) at face value, which 
may be unduly optimistic.}
\end{itemize}
\noindent

When including the likelihood for the muon anomalous magnetic
moment, $a_\mu$, we calculate it combining the experimental 
and the theoretical uncertainties as follows:
\beq
{\cal L}_{a_\mu}=\frac{1}{\sqrt{2 \pi} \sigma}
e^{-(a_\mu^{th}-a_\mu^{exp})^2/2 \sigma^2} \,,
\label{eq:likamu}
\eeq 
where $\sigma^2=\sigma_{exp}^2+\sigma_{th}^2$, with
$\sigma_{exp}$ taken from (\ref{delta}) and $\sigma_{th}^2$
from (\ref{eq:sigmath}), replacing $m_h$ by $a_\mu$.

As is well known, the discrepancy (\ref{delta}) would place significant
constraints on the CMSSM parameter space, favouring $\mu > 0$, though we
do consider both signs of $\mu$. In fact, we find that $\mu > 0$ is 
favoured somewhat, even with the `pessimistic' version (\ref{deltatau}) of 
the $g_\mu - 2$ constraint.

\subsection{Density of Cold Dark Matter}

As already mentioned, we identify the relic density of LSPs with
$\Omega_{CDM} h^2$. In addition to the CMSSM parameters, the calculation
of $\Omega_{CDM} h^2$ involves some parameters of the Standard Model that
are poorly known, such as $m_t$ and $m_b$. The default values and 
uncertainties we assume
for these parameters have been mentioned above. Here we stress that both
these parameters should be allowed to run with the effective scale $Q$ at
which they contribute to the calculation of the relic density, which is
typically $Q \simeq 2 m_\chi$. This effect is particularly important when
treating the rapid-annihilation channels due to $\chi \chi \to A, H \to
X\bar{X}$
annihilations, but is non-negligible also in other parts of the CMSSM
parameter space.

Specifically, the location of the rapid-annihilation funnel
due to $A,H$ Higgs-boson exchange, which
appears in the region where $m_A \simeq 2 m_\chi$, depends significantly
on the determination of $m_A$~\cite{LS}. For this
determination, the input value of the running $\overline{MS}$ mass 
of  $m_b$ is a crucial parameter, 
and the appearance of the funnels depends noticeably on 
$m_b$~\cite{ganis,ftuning}. On the other hand, the exact location
of the focus-point region~\cite{focus} 
(also known as the hyperbolic branch of radiative symmetry breaking 
\cite{hbrsb}) depends sensitively on $m_t$ \cite{rs,ftuning,eos2}, which
dictates the scale of radiative
electroweak symmetry breaking~\cite{LNM}.   
 
In calculating the likelihood of the CDM density, we follow a similar
procedure as for the anomalous magnetic moment of the muon in
(\ref{eq:sigmath}, \ref{eq:likamu}), again taking into
account the contribution the uncertainties in $m_{t,b}$. In this case, we
take the experimental uncertainty from WMAP \cite{wmap,reion} and the
theoretical uncertainty from (\ref{eq:sigmath}), replacing $m_h$ by
$\Omega_\chi h^2$. We will see that the theoretical uncertainty plays a
very significant role in our analysis.

\subsection{The Total Likelihood}

The total likelihood function is computed by combining all the components
described above: \beq {\cal L}_{tot} = {\cal L}_h \times {\cal
L}_{bs\gamma} \times {\cal L}_{\Omega_\chi h^2} \times {\cal L}_{a_\mu}
\eeq In what follows, we consider the CMSSM parameter space at fixed
values of $\tan \beta$ = 10, 35, and 50 with $A_0 = 0$. For $\tan \beta $
= 10 and 35, we compute the likelihood function for both signs of $\mu$,
but not for $\tan \beta = 50$, since in the case the choice $\mu < 0$ does
not provide a solution of the RGEs with radiative electroweak symmetry
breaking.

The likelihood function in the CMSSM can be considered as a function of
two variables, ${\cal L}_{tot}(m_{1/2},m_0)$, where $m_{1/2}$ and $m_0$
are the unified GUT-scale gaugino and scalar masses respectively. When
plotting confidence levels as iso-likelihood contours in the corresponding
$(m_{1/2}, m_0)$ planes, we normalize likelihood function by setting the
volume integral
\beq
 \int  {\cal L}_{tot} \, dm_0 \, dm_{1/2} \; = \; 1
\eeq
for each value of $\tan \beta$, combining where appropriate both signs of 
$\mu$. We also compare the 
integrals of the likelihood function over the coannihilation and 
focus-point regions, and for different values of $\tan \beta$.

For most of the results presented below, we perform the analysis over the
range $m_{1/2} = 100$~GeV up to 1000~GeV for $\tan \beta = 10$ and up to 2
TeV for $\tan \beta$ = 35 and 50.  The upper limit on $m_0$ is taken to be
the limit where solutions for radiative electroweak symmetry breaking are
possible and the range includes the focus-point region at large $m_0$. We
discuss below the sensitivity of our results to the choice of the upper
limit on $m_{1/2}$.

\section{Widths of Allowed Strips in the CMSSM Parameter Space}
\label{sec:strips}

We begin by first presenting the global likelihood function along cuts
through the $(m_{1/2}, m_0)$ plane, for different choices of $\tan \beta$,
the sign of $\mu$ and $m_{1/2}$. These exhibit the relative importance of
experimental errors and other uncertainties, as well as the potential
impact of the $g_\mu - 2$ measurement.

\begin{figure}
\begin{center}
\mbox{\epsfig{file=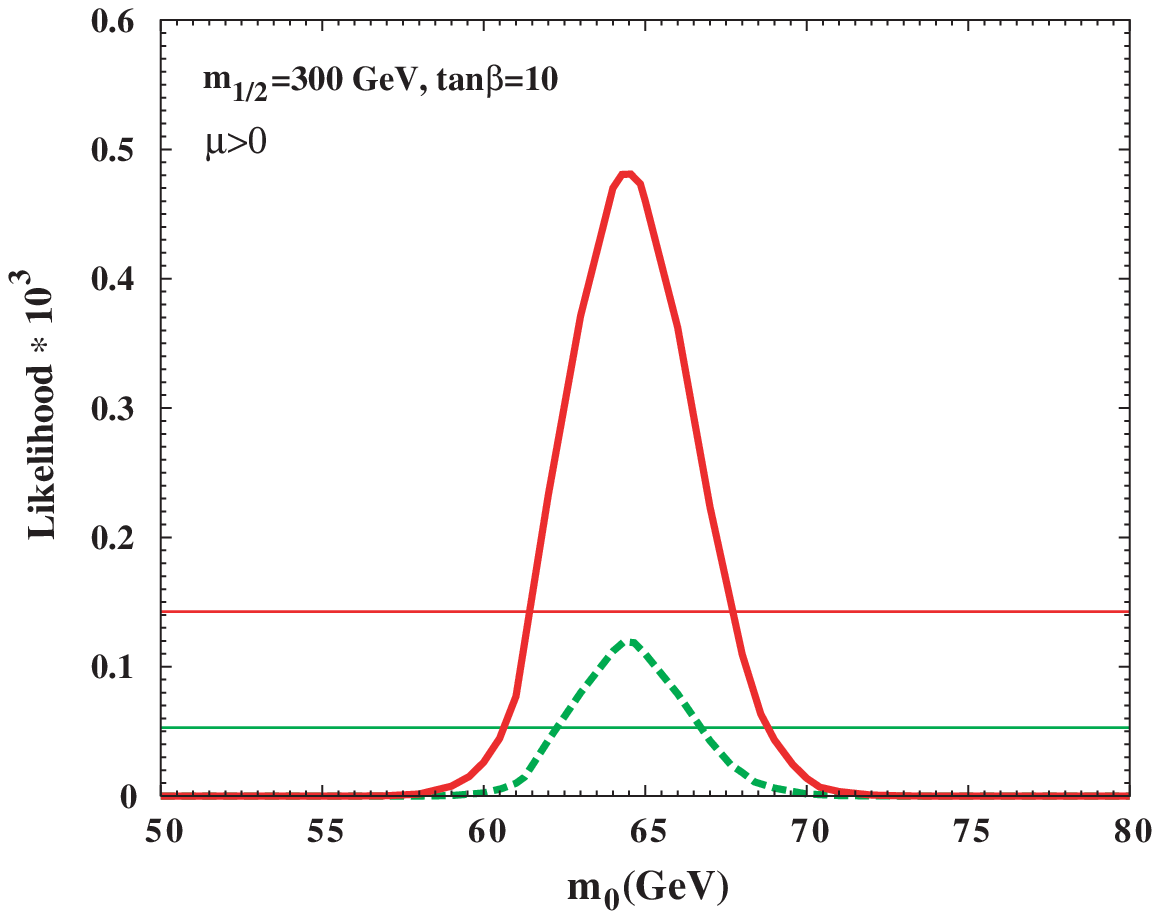,height=6cm}}
\mbox{\epsfig{file=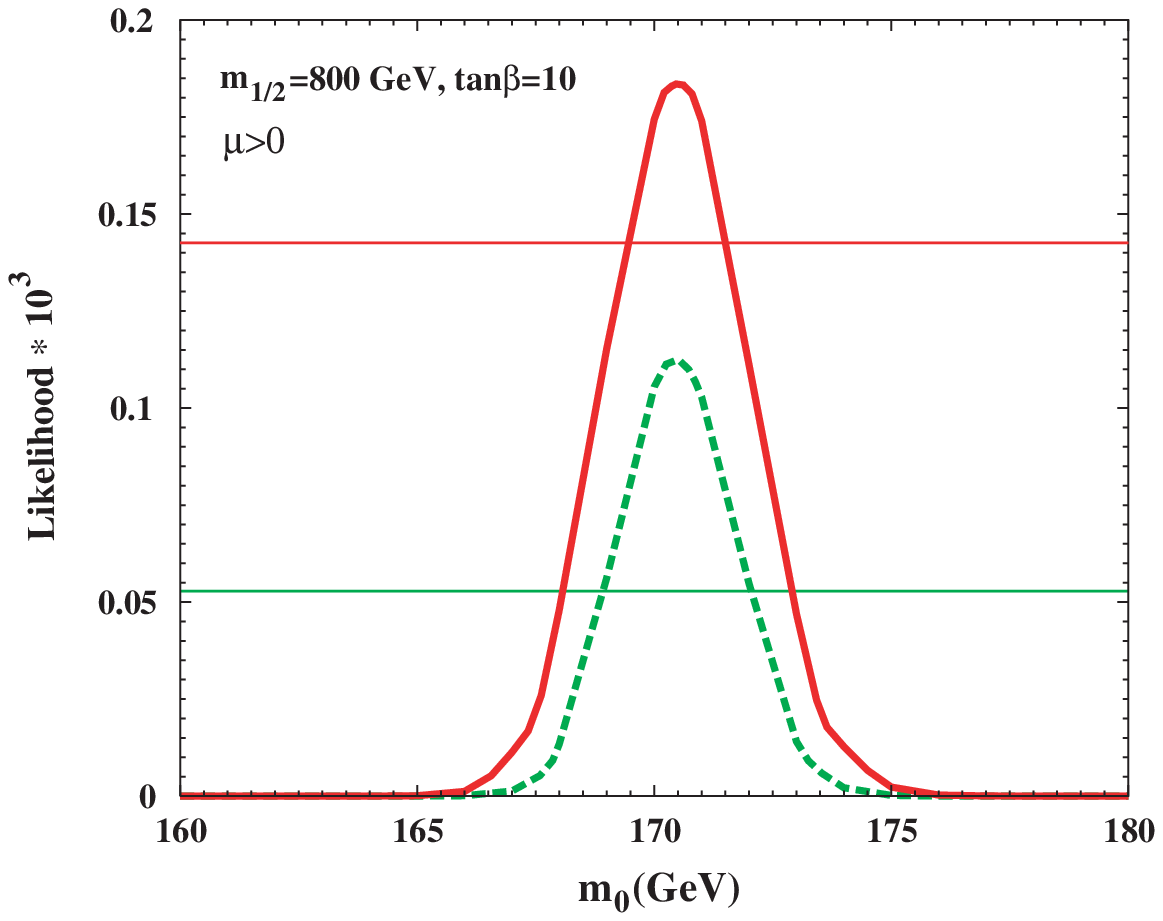,height=6cm}}
\end{center} 
\caption{\it
The likelihood function along slices in $m_0$ through the CMSSM parameter
space for $\tan \beta = 10, A_0 = 0, \mu> 0$ and $m_{1/2} = 300, 800$~GeV in the
left and right panels, respectively. The solid red curves show
the total likelihood function and the green dashed curve is the likelihood 
function with $\Delta m_t = \Delta m_b = 0$. Both analyses include the $g_\mu - 
2$ likelihood calculated using $e^+ e^-$ data. The horizontal  
lines show the 68\% confidence level of the likelihood function for each case.}
\label{fig:WMAP}
\end{figure}

We first display in Fig.~\ref{fig:WMAP} the likelihood along slices
through the CMSSM parameter space for $\tan \beta = 10, A_0 = 0, \mu > 0,$
and $m_{1/2} = 300$ and 800~GeV in the left and right panels,
respectively, plotting the likelihood as a function of $m_0$ in the
neighborhood of the coannihilation region \cite{coann}. The solid red curves show the
total likelihood function calculated including the uncertainties which
stem from the experimental errors in $m_t$ and $m_b$.  The green dashed
curves show the likelihood calculated without these uncertainties, i.e.,
we set $\Delta m_t = \Delta m_b = 0$. We see that these errors have
significant effects on the likelihood function.  In each panel, the
horizontal lines correspond to the 68\% confidence level of the respective
likelihood function.  The likelihood functions shown here include ${\cal
L}_{a_\mu}$ calculated using $e^+ e^-$ data. 
For these values of $m_{1/2}$
and $m_0$ with $\mu > 0$, the constraint from $g_\mu - 2$ is not very
significant. 
For reference, we present in Table~\ref{table:cl} and \ref{table:cl2}
the values of the likelihood functions corresponding to the 68\%, 90\%,
and 95\% CLs for each choice of $\tan \beta$ and $\Delta m_t$.

\begin{table}
\begin{center}
\caption{\it Likelihood values ($\times 10^3$), including $g_\mu -2$, for the
68\%, 90\%, 
and 95\% CLs for different choices of $\tan \beta$ and the uncertainty 
in $m_t$.}
\vspace*{3mm}
\label{table:cl}
\begin{tabular}{|c|c|c|c|c|c|}
\hline \hline
$\tan\beta$ & CL  & $\Delta m_t =  5$ GeV   &    $\Delta m_t =  1$ GeV 
                     & $\Delta m_t =  0.5$ GeV & $\Delta m_t = 0$ GeV \\ 
\hline
   & 68\%  & 0.14                 & 0.13                  & 0.087                 & 0.046 \\ 
10 & 90\%  & 3.0 $\times 10^{-3}$ & 1.4 $\times 10^{-4}$  & 2.5 $\times 10^{-4}$  & 0.021 \\
   & 95\%  & 2.9 $\times 10^{-5}$ & 6.2 $\times 10^{-5}$  &  1.1 $\times 10^{-4}$ & 0.011 \\
\hline
   & 68\%  & 2.2 $\times 10^{-4}$ & 2.0 $\times 10^{-4}$  & 2.7 $\times 10^{-4}$ 
                                                                     & 1.1$\times 10^{-3}$   \\ 
35 & 90\%  & 2.8 $\times 10^{-5}$ & 5.0 $\times 10^{-5}$  & 7.5 $\times 10^{-5}$ 
                                                                     & 1.8$\times 10^{-4}$  \\
   & 95\%  & 1.1 $\times 10^{-5}$ & 2.7 $\times 10^{-5}$  & 3.9 $\times 10^{-5}$  
                                                                     & 6.8$\times 10^{-5}$  \\
\hline
   & 68\%  & 5.3 $\times 10^{-4}$ & 5.7 $\times 10^{-4}$  & 5.4 $\times 10^{-4}$  
                                                                     & 7.0  $\times 10^{-4}$\\ 
50 & 90\%  & 8.7 $\times 10^{-5}$ & 7.7 $\times 10^{-5}$  & 1.0 $\times 10^{-4}$  
                                                                     & 1.9  $\times 10^{-4}$ \\
   & 95\%  & 2.3 $\times 10^{-5}$ & 3.2 $\times 10^{-5}$  & 4.6 $\times 10^{-5}$  
                                                                     & 8.2 $\times 10^{-5}$ \\
\hline
\hline
\end{tabular}
\end{center}
\end{table}

\begin{table}
\begin{center}
\caption{\it Likelihood values ($\times 10^3$), excluding $g_\mu -2$, for the
68\%, 90\%, 
and 95\% CLs for different choices of $\tan \beta$ and the uncertainty 
in $m_t$.}
\vspace*{3mm}
\label{table:cl2}
\begin{tabular}{|c|c|c|c|c|c|}
\hline \hline
$\tan\beta$ & CL  & $\Delta m_t =  5$ GeV   &    $\Delta m_t =  1$ GeV 
                     & $\Delta m_t =  0.5$ GeV & $\Delta m_t = 0$ GeV \\ 
\hline
   & 68\%  & 0.059                & 1.6 $\times 10^{-3}$  & 9.6 $\times 10^{-4}$  
                                                                              & 0.052   \\ 
10 & 90\%  & 5.6 $\times 10^{-5}$ & 1.3 $\times 10^{-4}$  & 2.4 $\times 10^{-4}$  
                                                                              & 0.024   \\
   & 95\%  & 4.0 $\times 10^{-5}$ & 7.9 $\times 10^{-5}$  & 1.3 $\times 10^{-4}$  
                                                                              & 0.011  \\
\hline
   & 68\%  & 2.4 $\times 10^{-4}$ & 1.9 $\times 10^{-4}$  & 2.4 $\times 10^{-4}$  
                                                              & 7.8$\times 10^{-4}$   \\ 
35 & 90\%  & 2.3 $\times 10^{-5}$ & 5.0 $\times 10^{-5}$  & 7.7 $\times 10^{-5}$  
                                                              & 1.6$\times 10^{-4}$  \\
   & 95\%  & 1.2 $\times 10^{-5}$ & 2.8 $\times 10^{-5}$  & 4.0 $\times 10^{-5}$  
                                                              & 7.5$\times 10^{-5}$  \\
\hline
   & 68\%  & 3.3 $\times 10^{-4}$ & 3.5 $\times 10^{-4}$  & 4.3 $\times 10^{-4}$  
                                                              & 6.4  $\times 10^{-4}$\\ 
50 & 90\%  & 4.2 $\times 10^{-5}$ & 6.4 $\times 10^{-5}$  & 1.0 $\times 10^{-4}$  
                                                              & 1.8  $\times 10^{-4}$ \\
   & 95\%  & 2.0 $\times 10^{-5}$ & 3.4 $\times 10^{-5}$  & 5.0 $\times 10^{-5}$  
                                                               & 8.5 $\times 10^{-5}$ \\
\hline
\hline
\end{tabular}
\end{center}
\end{table}

\begin{figure}
\begin{center}
\mbox{\epsfig{file=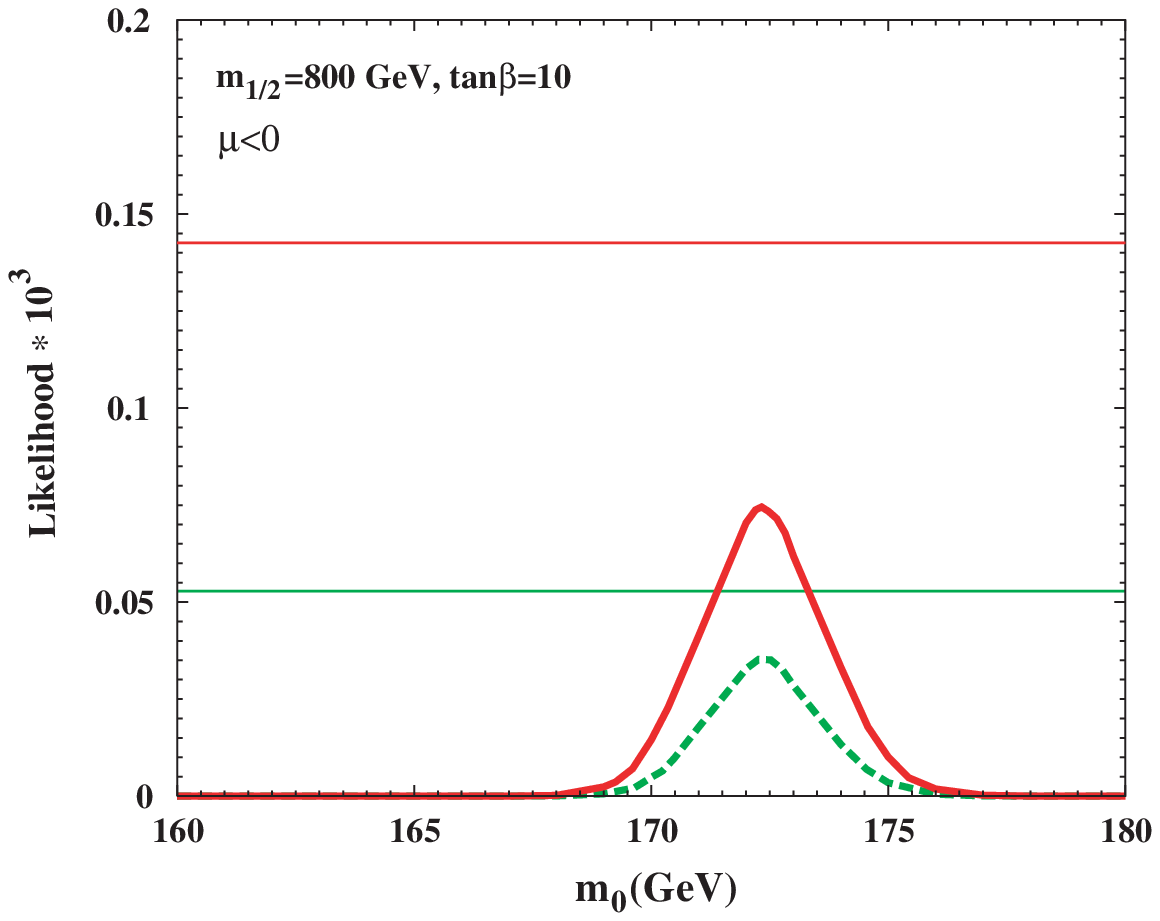,height=6cm}}
\mbox{\epsfig{file=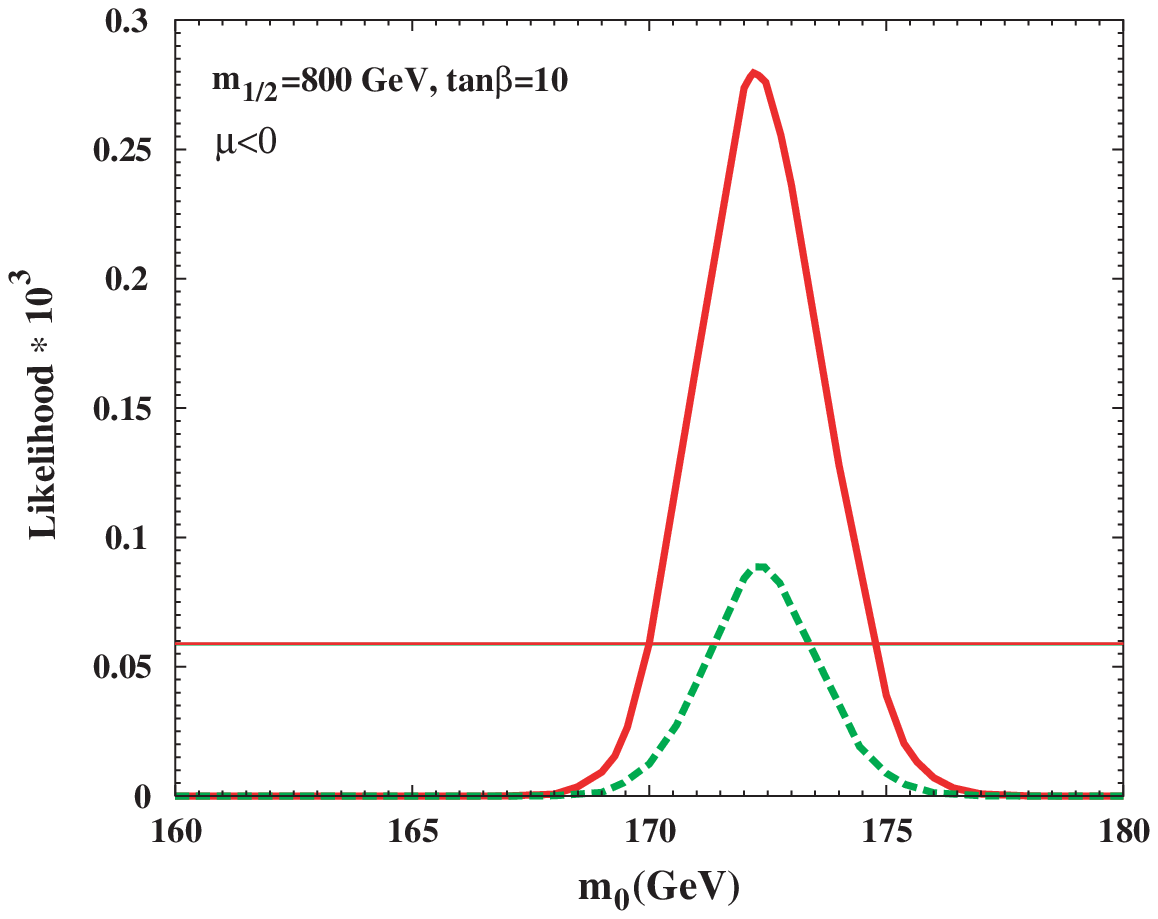,height=6cm}}
\end{center} 
\caption{\it
As in Fig.~\protect\ref{fig:WMAP} for $\tan \beta = 10, A_0 = 0, \mu< 0$
and $m_{1/2} = 800$~GeV. The $g_\mu-2$ constraint is included (excluded) in the
left (right) panels. In the right panel the 68\% CLs for both cases are
incidentally closed to each other. }
\label{fig:WMAPn}
\end{figure}

When $\mu < 0$, the $g_\mu-2$ information plays a more important role, as
exemplified in Fig.~\ref{fig:WMAPn}, where we show the likelihood in the
coannihilation region for $m_{1/2} = 800$ GeV. For $m_{1/2} = 300$ GeV,
the likelihood is severely suppressed (see the discussion below) and we do
not show it here.

We now discuss the components of the likelihood function which affect the
relative heights along the peaks shown in Fig.~\ref{fig:WMAP}. In the case
$m_{1/2} = 300$~GeV, the likelihood increases when the errors in $m_t$ and
$m_b$ are included, due to two dominant effects.  1) The total integrated
likelihood is decreased when the errors are turned on (by a factor of $\sim 2$
when $g_\mu-2$ is included and by a factor of $\sim 3$ when it is omitted, 
for $\tan
\beta = 10$), 
so the normalization constant, ${\cal N}$, becomes larger, and 2) since $m_{1/2}
= 300$~GeV corresponds to the lower limit on $m_{1/2}$ due to the
experimental bound on the Higgs mass, the Higgs contribution to the
likelihood increases when the uncertainties in the heavy quark masses are
included. When $m_{1/2} = 800$~GeV, 
it is primarily the normalization effect which results in
an overall increase. The Higgs mass contribution at this value of $m_{1/2}$ is
essentially ${\cal L}_{h_{exp}} = 1$.
We remind the reader that the value of the
likelihood itself has no meaning.  Only the relative likelihoods (for a
given normalization) carry any statistical information, which is
conveyed here partially by the comparison to the respective 68\% CL
likelihood values.

\begin{figure}
\begin{center}
\mbox{\epsfig{file=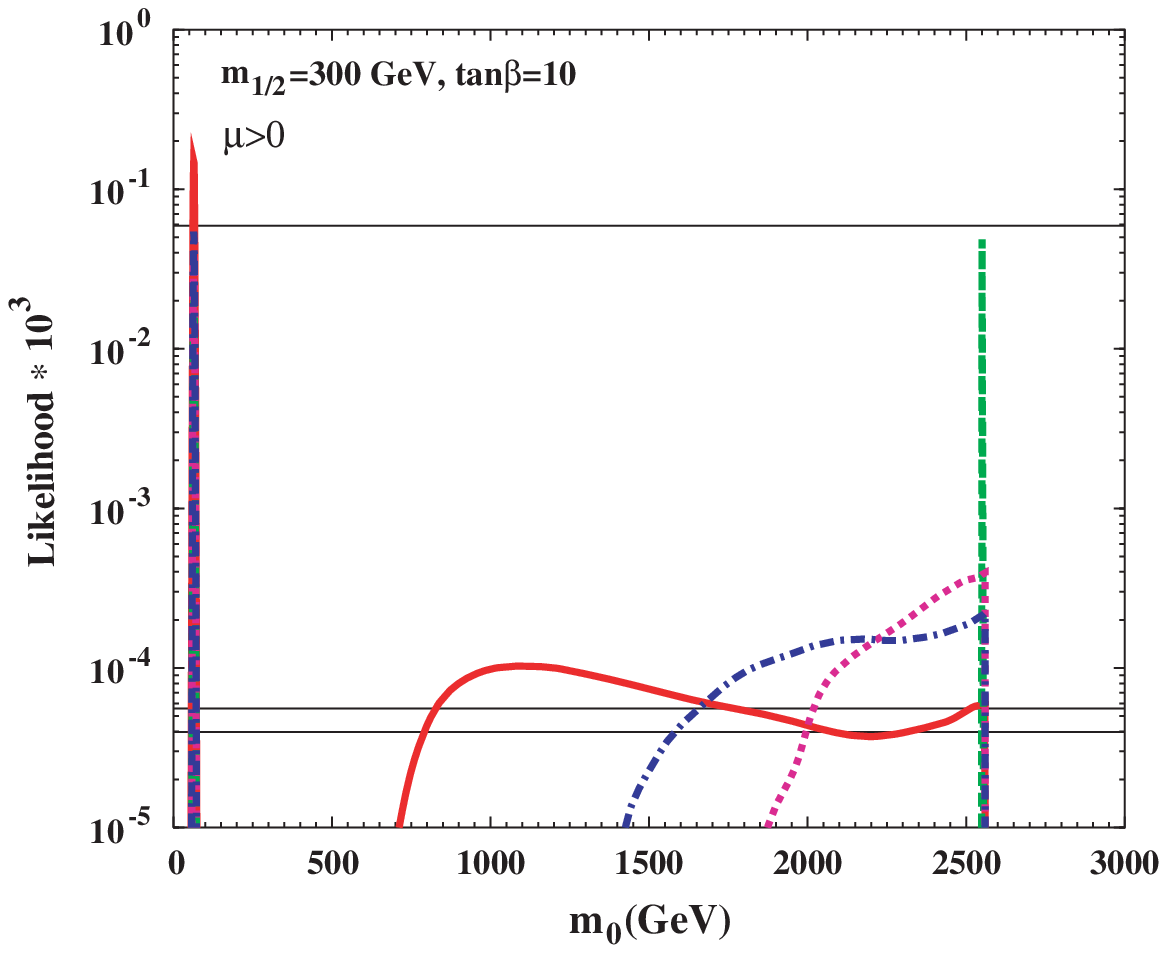,height=6cm}}
\mbox{\epsfig{file=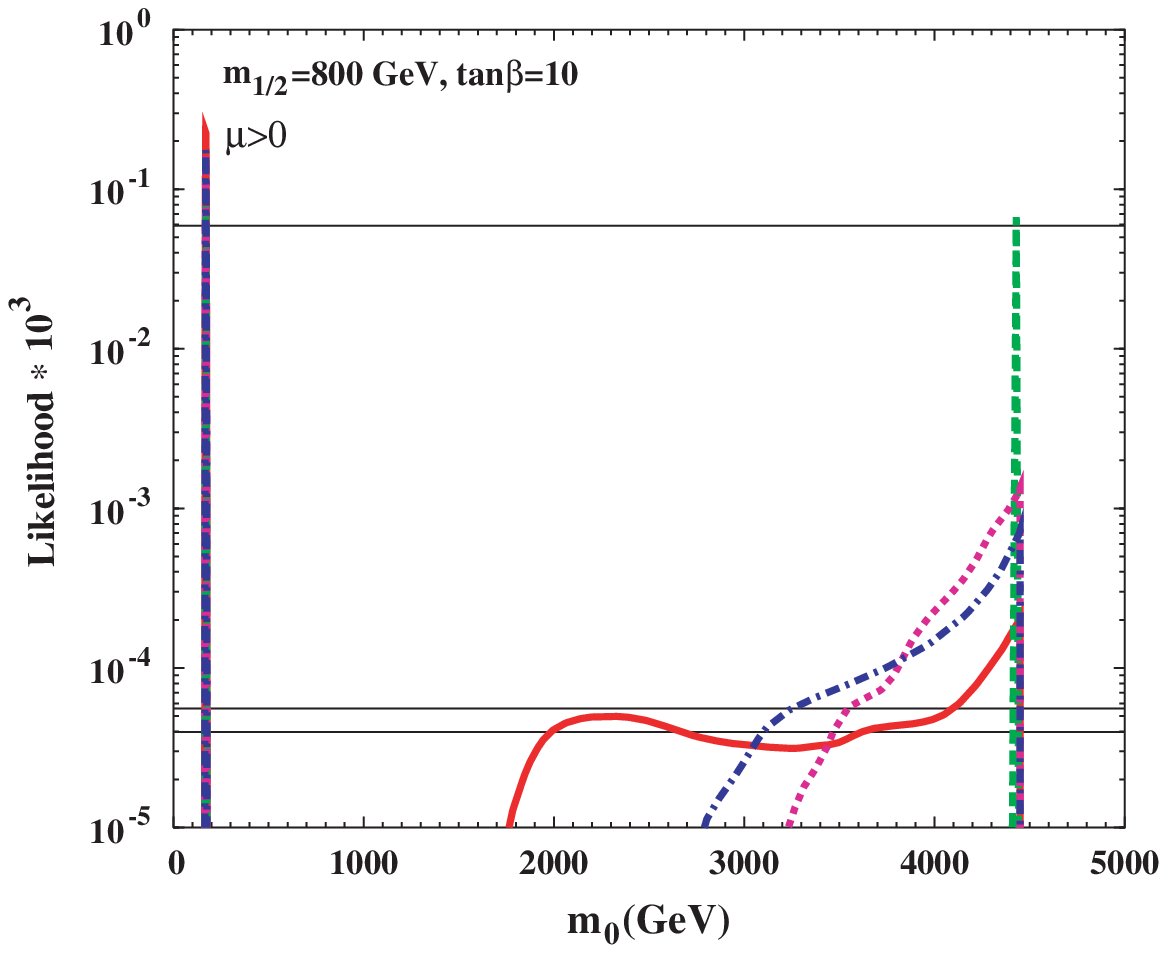,height=6cm}}
\end{center}   
\begin{center}
\mbox{\epsfig{file=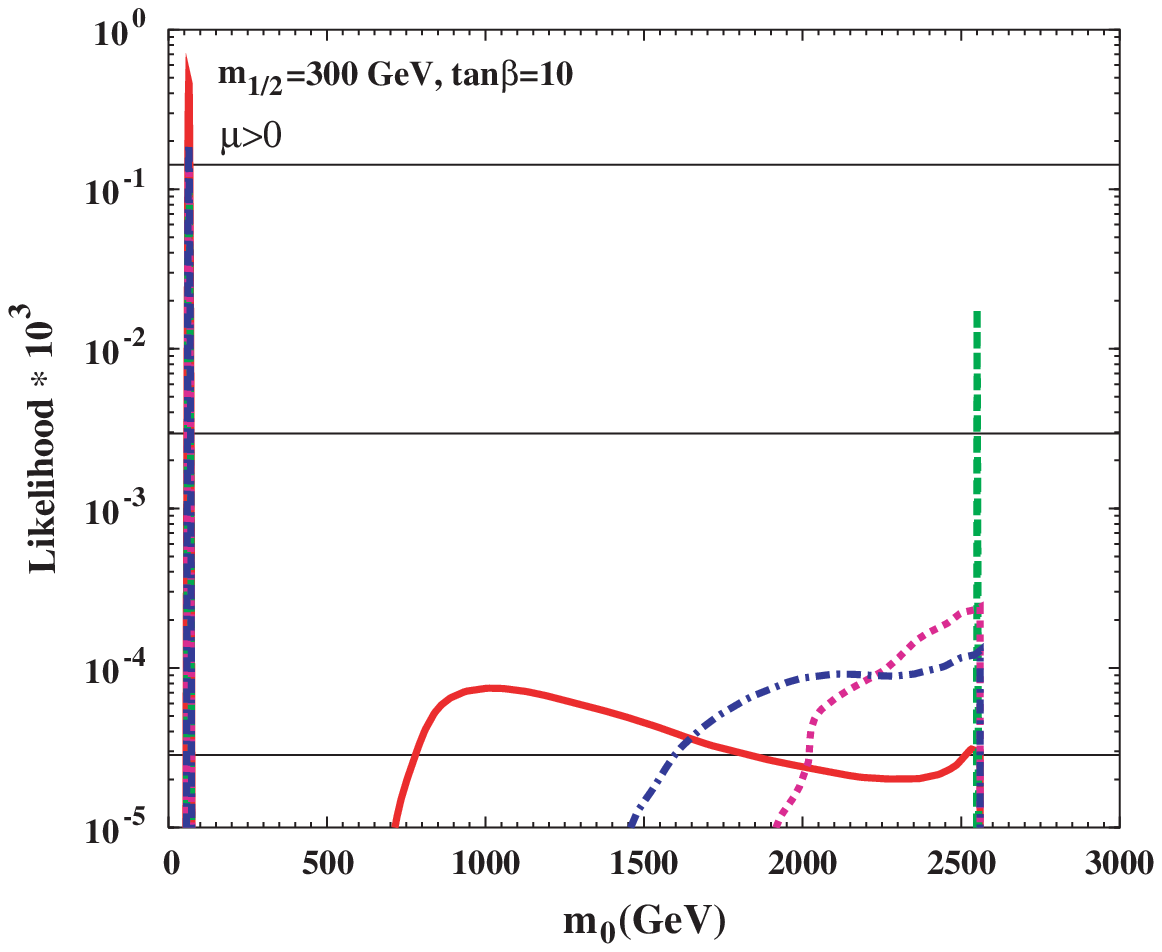,height=6cm}}
\mbox{\epsfig{file=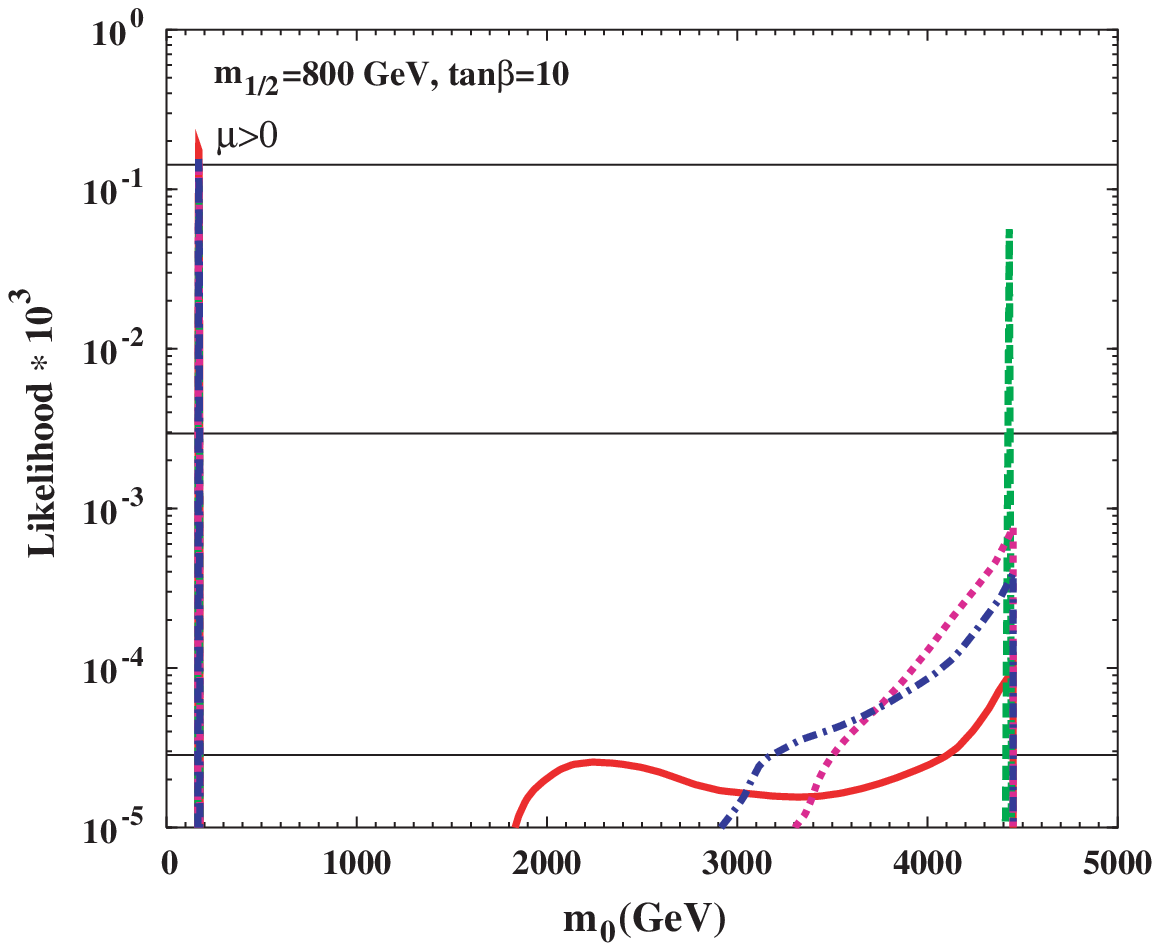,height=6cm}}
\end{center} 
\caption{\it
As in Fig.~\ref{fig:WMAP}, but for slices at fixed $m_{1/2}$ that 
include also the focus-point region at large $m_0$.
The red (solid) curves are calculated using the current errors in 
$m_t$ and $m_b$, the green dashed curve with no error in $m_t$,
the violet dotted lines with $\Delta m_t = 0.5$~GeV, and
the blue dashed-dotted lines with $\Delta m_t = 1$~GeV. 
In the upper two figures, the $g_\mu-2$ constraint has not been applied.
}
\label{fig:WMAPFP}
\end{figure}

In Fig.~\ref{fig:WMAPFP}, we extend the previous slices through the CMSSM
parameter space to the focus-point region at large $m_0$.  The solid (red)
curve corresponds to the same likelihood function shown by the solid (red)
curve in Fig.~\ref{fig:WMAP}, and the peak at low $m_0$ is due to the
coannihilation region. The peak at $m_0 \simeq 2500 (4500)$ GeV for
$m_{1/2} = 300 (800)$ GeV is due to the focus-point region~\footnote{We
should, in
addition, point out that different locations for the focus-point region
are found in different theoretical codes, pointing to further systematic
errors that are currently not quantifiable.}.  The $g_\mu -
2$ constraint is not taken into account in the upper two figures of this
panel. Also shown in Fig. \ref{fig:WMAPFP} are the 68\%, 90\%, and 95\% CL
lines, corresponding to the iso-likelihood values of the fully integrated
likelihood function corresponding to the solid (red) curve. As one can
see, one of the effects of the $g_\mu - 2$ constraint (even at its
recently reduced significance) is a suppression of the likelihood function
in the focus-point region. 

\begin{figure}
\begin{center}
\mbox{\epsfig{file=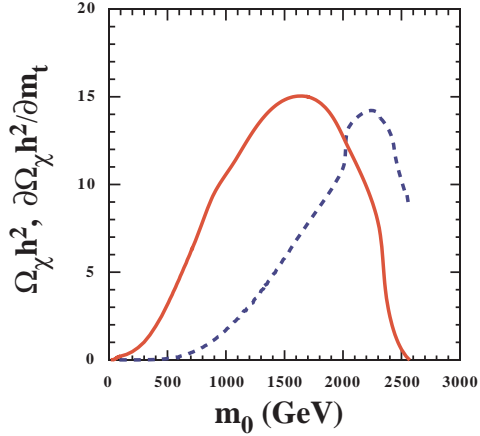,height=6cm}}
\end{center} 
\caption{\it
The value of $\Omega_\chi h^2$ (solid) and $\partial \Omega h^2/\partial m_t$ 
(dashed) as functions of $m_0$ for $\tan \beta = 10, A_0 = 0, \mu> 0$ and 
$m_{1/2} = 300$~GeV, corresponding to the slice shown in 
Fig.~\protect\ref{fig:WMAPFP}c.  } 
\label{fig:oh2sig}
\end{figure}

The focus-point peak is suppressed relative to the coannihilation peak at
low $m_0$ because of the theoretical sensitivity to the experimental
uncertainty in the top mass.  We recall that the likelihood function is
proportional to $\sigma^{-1}$, and that $\sigma$ which scales with
$\partial (\Omega_\chi h^2 )/ \partial m_t$,  is very large at large
$m_0$~\cite{ftuning}. This sensitivity is shown in Fig.~\ref{fig:oh2sig},
which plots both $\Omega_\chi h^2$ and $\partial (\Omega_\chi h^2 )/
\partial m_t$ for the cut corresponding to Fig.~\ref{fig:WMAPFP}c. Notice
that, for the two values of $m_0$ with $\Omega_\chi h^2 \sim 0.1$,
corresponding to the coannihilation and focus-point regions, the error due
to the uncertainty in $m_t$ is far greater in the focus-point region than
in the coannihilation region.  Thus, even though the exponential in ${\cal
L}_{\Omega_\chi h^2}$ is of order unity near the focus-point region when
$\Omega_\chi h^2 \simeq 0.1$, the prefactor is very small due the large
uncertainty in the top mass. This accounts for the factor of $\ga 1000$ suppression
seen in Fig.~\ref{fig:WMAPFP} when comparing the two peaks of the solid red 
curves.

We note also that there is another broad, low-lying peak at intermediate
values of $m_0$. This is due to a combination of the effects of $\sigma$
in the prefactor and the exponential.  We expect a bump to occur when the
Gaussian exponential is of order unity, i.e., $\Omega_\chi h^2 \sim
\sqrt{2}\Delta m_t \, \partial \Omega_\chi h^2/\partial m_t$.  From the solid
curve in Fig.~\ref{fig:oh2sig}, we see that $\Omega_\chi h^2 \sim 10$ at
large $m_0$ for our nominal value $m_t$ = 175 GeV, but it varies 
significantly as one samples the favoured range of $m_t$ within its 
present uncertainty.
The competition between the exponential and the prefactor
would require a large theoretical uncertainty in $\Omega_\chi h^2$: 
$\partial \Omega_\chi h^2/\partial m_t \sim 2$ for $\Delta m_t =
5$ GeV.  From the dashed curve in Fig.~\ref{fig:oh2sig}, we see that this
occurs when $m_0 \sim 1000$ GeV, which is the position of the broad
secondary peak in Fig.~\ref{fig:WMAPFP}a. At higher $m_0$, $\sigma$
continues to grow, and the prefactor suppresses the likelihood function
until $\Omega_\chi h^2$ drops to $\sim 0.1$ in the focus-point region.

As is clear from the above discussion, the impact of the present
experimental error in $m_t$ is particularly important in this region. This
point is further demonstrated by the differences between the curves in
each panel, where we decrease {\it ad hoc} the experimental uncertainty in
$m_t$.  As $\Delta m_t$ is decreased, the intermediate bump blends into
the broad focus-point peak.  
Once again, this can be understood from Fig.~\ref{fig:oh2sig}, where 
we see that as $\Delta m_t$  is decreased, we require a
large sensitivity to $m_t$ in order to get an increase in ${\cal L}$.
This happens at higher $m_0$, and thus explains the shift in the
intermediate bump to higher $m_0$ as $\Delta m_t$ decreases.
When the uncertainties in $m_t$ and $m_b$ are
set to 0, we obtain a narrow peak in the focus-point region. This is
suppressed relative to the coannihilation peak, due to the effect of the
$g_\mu - 2$ contribution to the likelihood.  

We can now understand better Tables~\ref{table:cl} and
\ref{table:cl2} for $\tan \beta = 10$. For the cases with $\Delta
m_t \neq 0$ in Table~\ref{table:cl} and $\Delta m_t = 5$~GeV in
Table~\ref{table:cl2}, the coannihilation peak is much higher than the
focus-point peak, so that the 68\% CL (or even the 80\% CL) does not
include the focus point. To reach the 90\% CL, we need to include some
part of the focus point, and this explains why the 68\% CL is much higher
than the 90\% CL. The $\Delta m_t = 1$~GeV case in Table~\ref{table:cl2}
is a peculiar one in which the integral over the coannihilation peak is
already around 68\% of the total integral and, because the focus point
peak is flat and broad, we do not need to change the level much to get the
90\% CL. In the cases with $\Delta m_t = 0$, and also $\Delta m_t =
0.5$~GeV in Table~\ref{table:cl2}, the focus-point peak is also relatively
high and already contributes at the 68\% CL. Therefore we do not see an
order of magnitude change between the 68\% CL and the 90\%~CL.

\begin{figure}
\begin{center}
\mbox{\epsfig{file=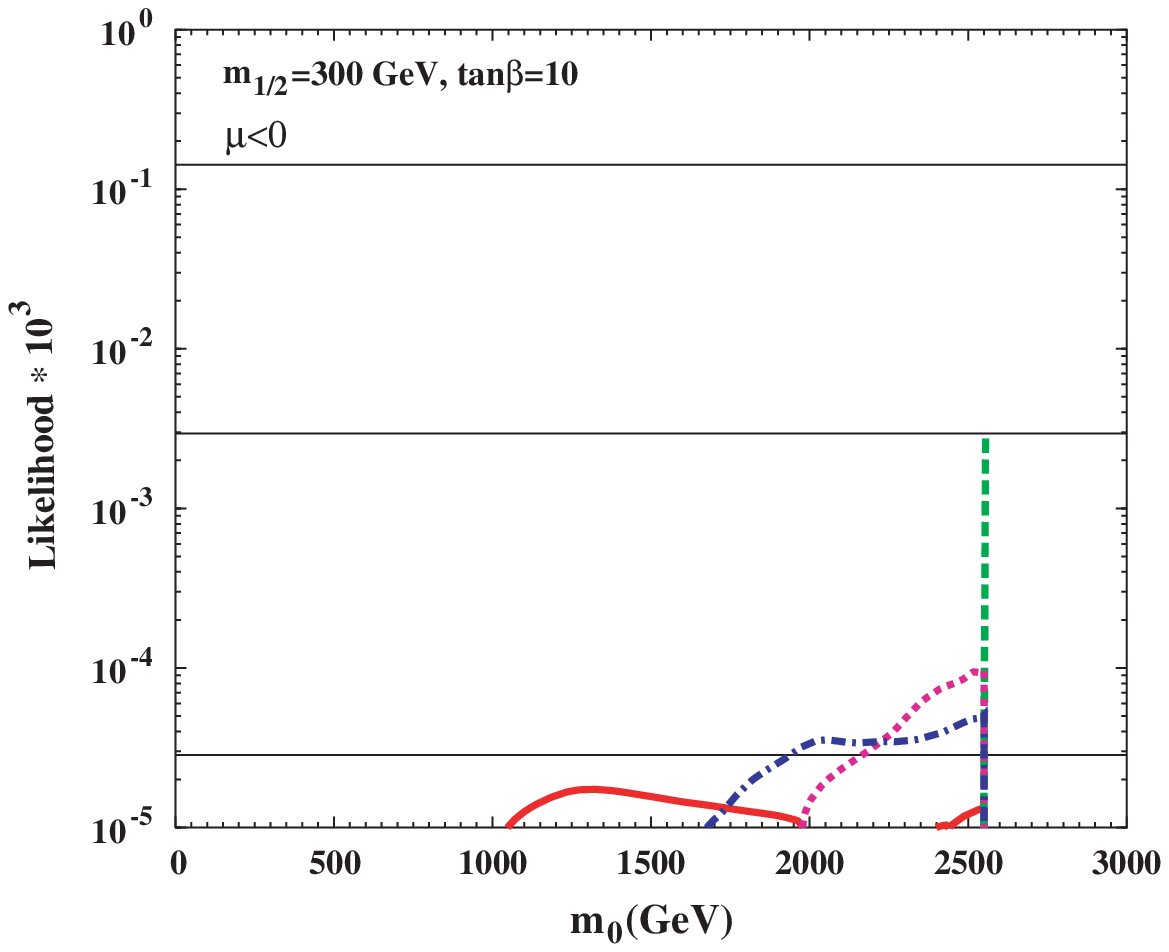,height=6cm}}
\mbox{\epsfig{file=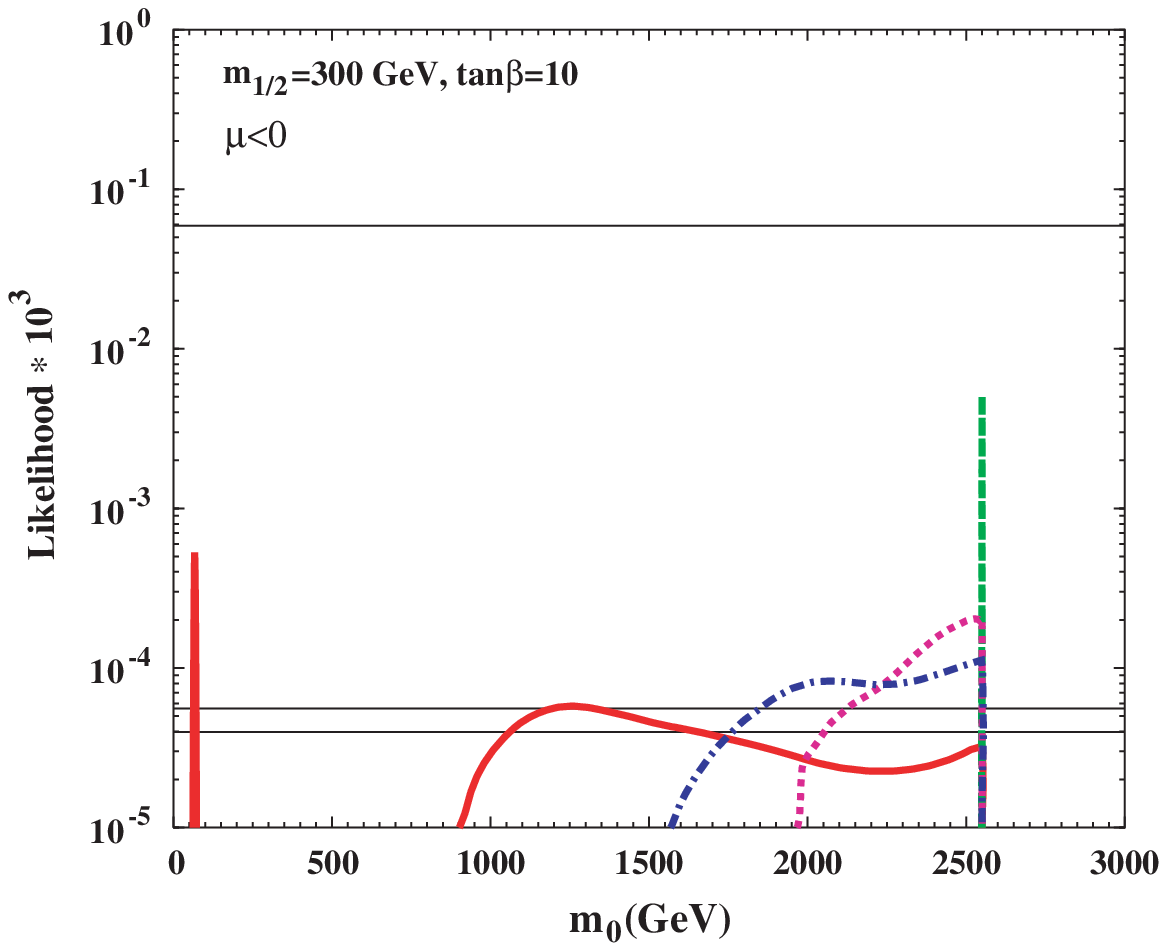,height=6cm}}
\end{center} 
\caption{\it
As in Fig.~\ref{fig:WMAPFP}, but for $\mu < 0$ and $m_{1/2} = 300$~GeV, 
including (excluding) the $g_\mu - 2$ contribution to the global 
likelihood in the left (right) panel.
}
\label{fig:WMAPFPn}
\end{figure}

As one would expect, the effect of the $g_\mu - 2$ constraint is more
pronounced when $\mu < 0$.  This is seen in Fig.~\ref{fig:WMAPFPn} for the
cut with $m_{1/2} = 300$ GeV.  The most startling feature is the absence
of the coannihilation peak at low $m_0$ when the $g_\mu - 2$ constraint is
applied.  In this case, the focus-point region survives, because the
sparticle masses there are large enough for the supersymmetric
contribution to $g_\mu - 2$ to be acceptably small. The broad plateau at
intermediate $m_0$ is suppressed in this case, and the likelihood does not
reach the 95\% CL when $\Delta m_t = 5$ GeV. Another effect of the Higgs
mass likelihood can be seen by comparing the coannihilation regions for
the two signs of $\mu$ when $m_{1/2} = 300$ GeV and the $g_\mu - 2$
constraint is not applied. Because the Higgs mass constraint is stronger
when $\mu < 0$~\footnote{For the same Higgs mass $m_h$, one needs to go to
a higher value of $m_{1/2}$ when $\mu < 0$. For the choices $\tan \beta =
10$ and $m_{1/2} = 300$ GeV, we find using {\tt FeynHiggs} nominal values 
$m_h = 114.1$ GeV for
$\mu > 0 $ and $m_h = 112.8$ GeV when $\mu < 0$.}, the coannihilation peak
is suppressed when $\mu < 0$ relative to its height when $\mu > 0$.  We
note that part of the suppression here is due to the $b \to s \gamma$
constraint, which also favours positive $\mu$.

\begin{figure}
\begin{center}
\mbox{\epsfig{file=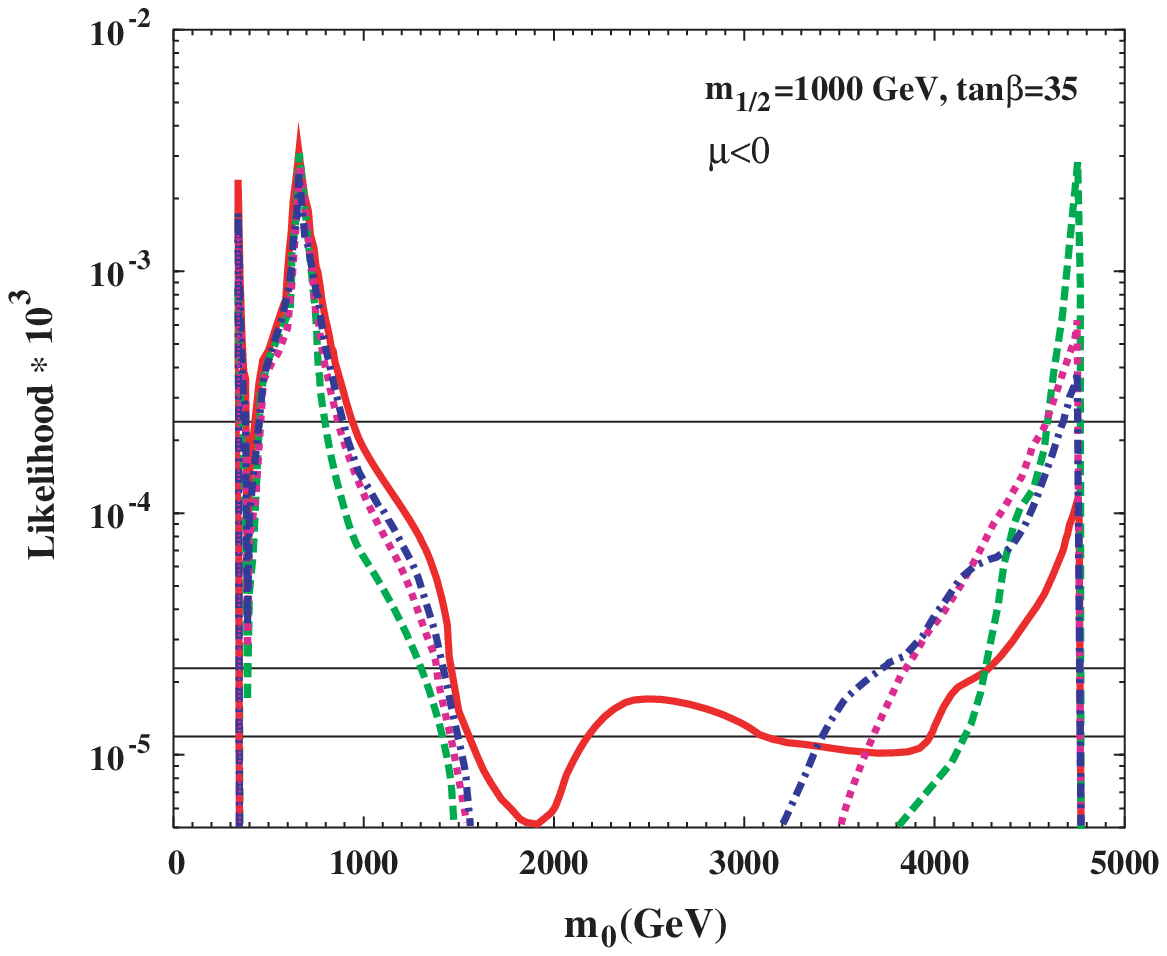,height=6cm}}
\mbox{\epsfig{file=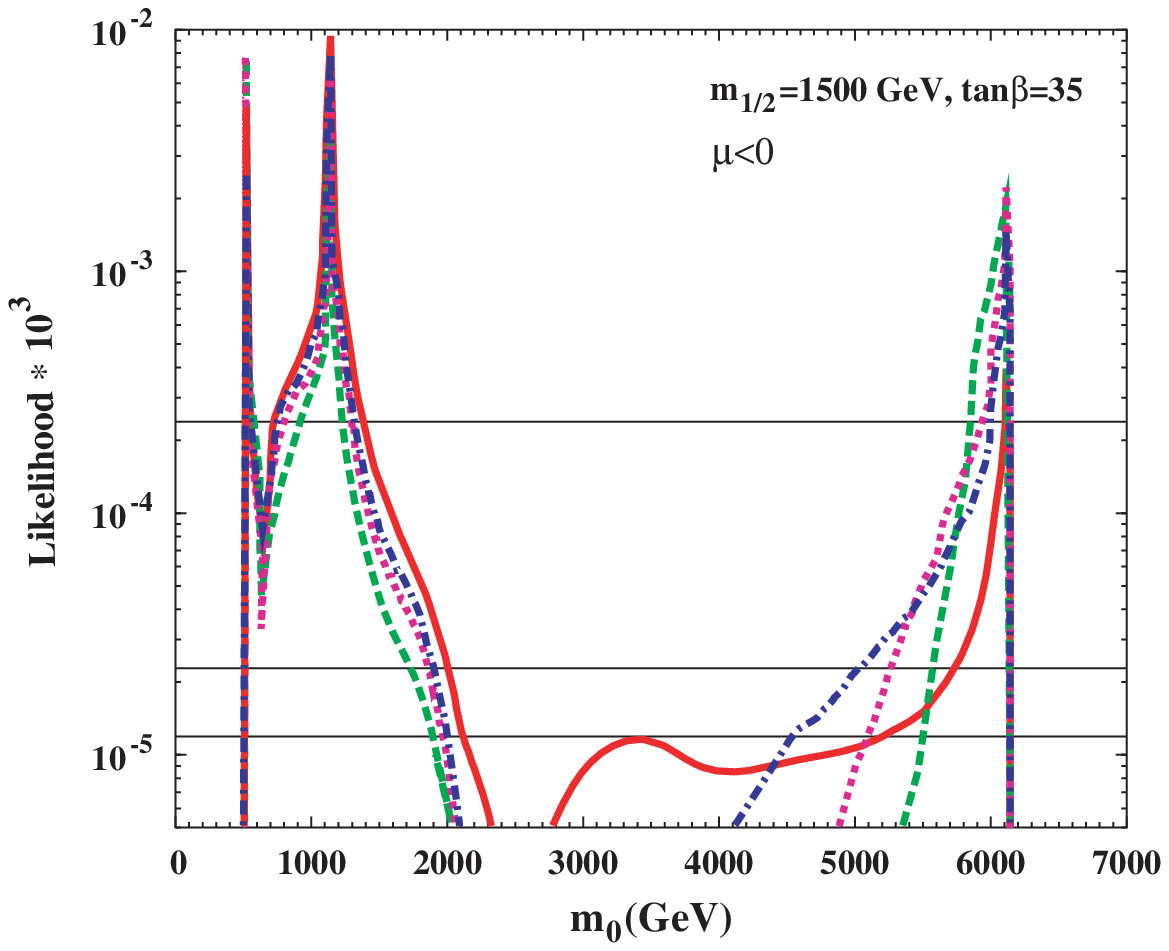,height=6cm}}
\end{center}   
\begin{center}
\mbox{\epsfig{file=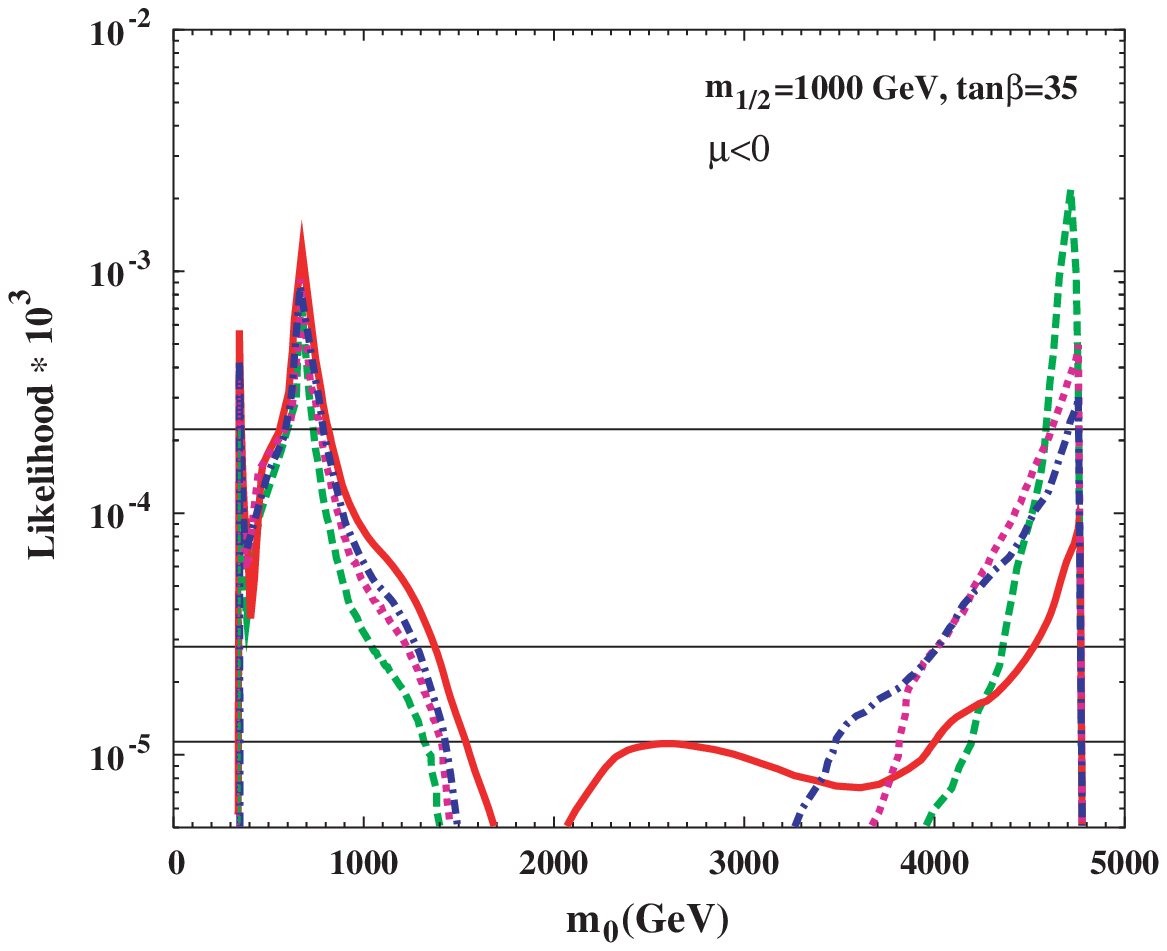,height=6cm}}
\mbox{\epsfig{file=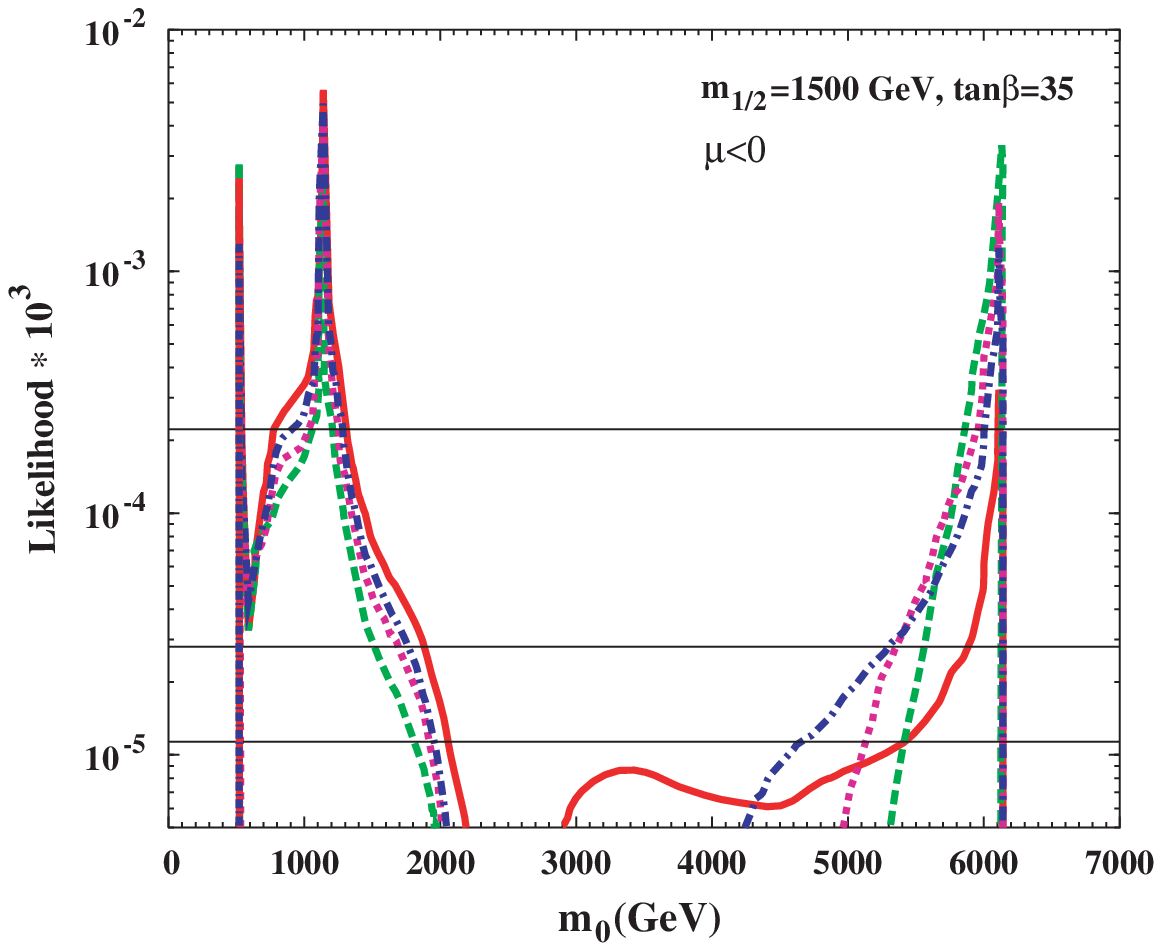,height=6cm}}
\end{center} 
\caption{\it
As in Fig.~\protect\ref{fig:WMAPFP} for $\tan \beta = 35, A_0 = 0, \mu< 0$ 
and $m_{1/2} = 1000, 1500$~GeV in the left and right panels.
The $g_\mu-2$ constraint is included (excluded) in the
bottom (top)  panels. }
\label{fig:WMAP35}
\end{figure}

We show in Fig.~\ref{fig:WMAP35} the likelihood function along cuts in the
$(m_{1/2}, m_0)$ plane for $\tan \beta = 35, A_0 = 0, \mu < 0$ and 
$m_{1/2} = 1000$~GeV (left panels) and $1500$~GeV (right panels). The 
$g_\mu - 2$ contribution to the likelihood is included in the bottom panels, 
but not in the top panels. The line styles are the same as in 
Fig.~\ref{fig:WMAPFP}, and we note that the behaviours in the focus-point 
regions are qualitatively similar. However, at $m_0 \sim 1000$~GeV the 
likelihood function exhibits double-peak structures reflecting the 
locations of the coannihilation strip and the rapid-annihilation funnels, 
whose widths depend on the assumed error in $m_t$, as can be seen by 
comparing the different line styles.

\begin{figure}
\begin{center}
\mbox{\epsfig{file=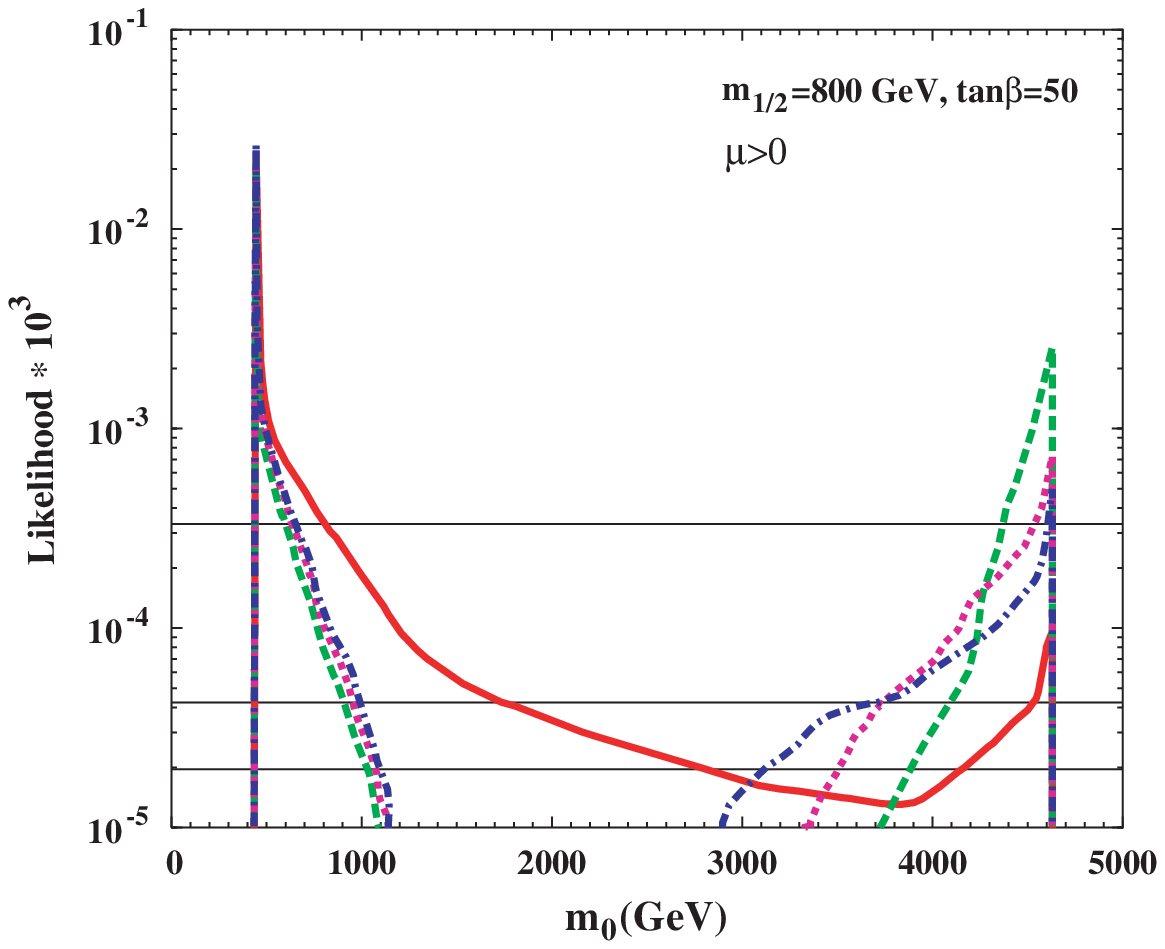,height=6cm}}
\mbox{\epsfig{file=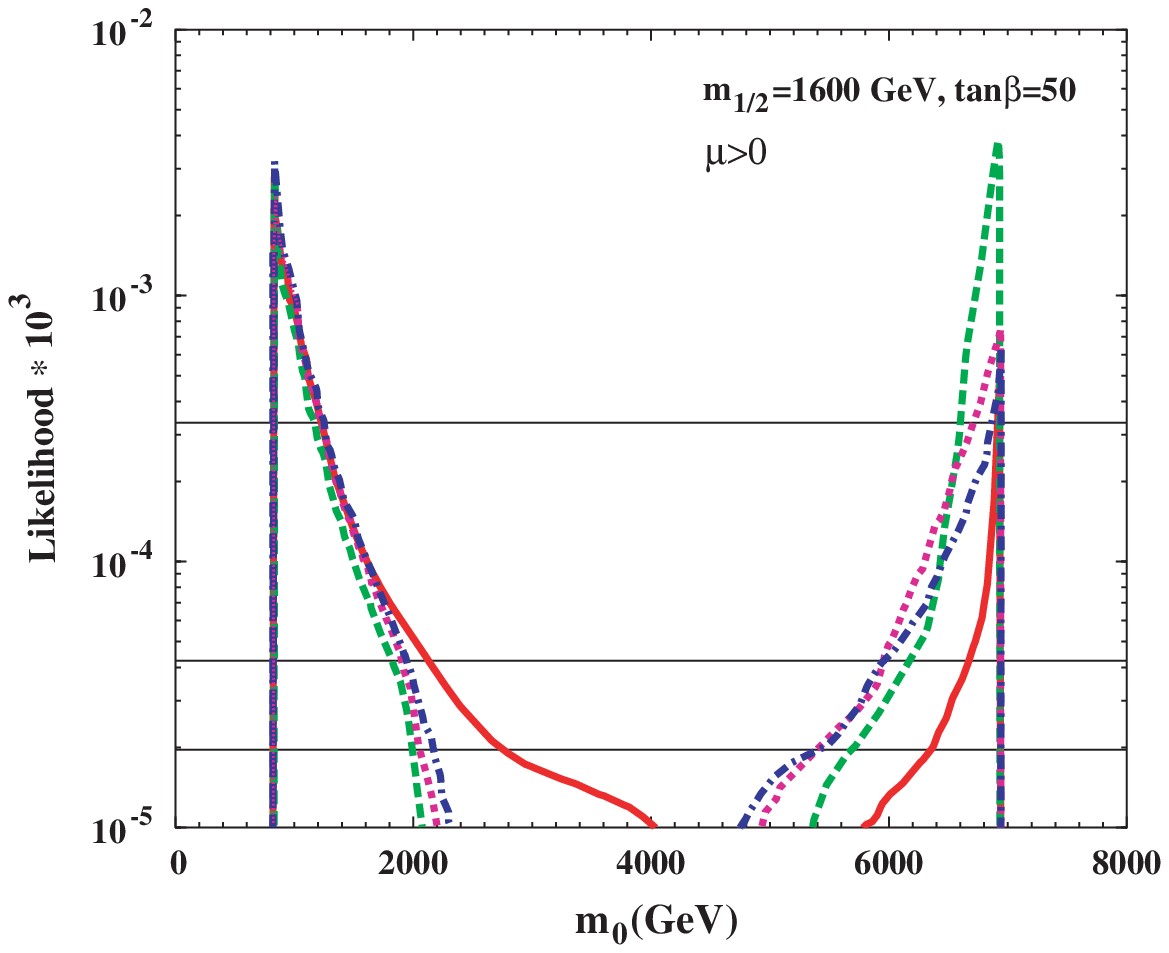,height=6cm}}
\end{center}   
\begin{center}
\mbox{\epsfig{file=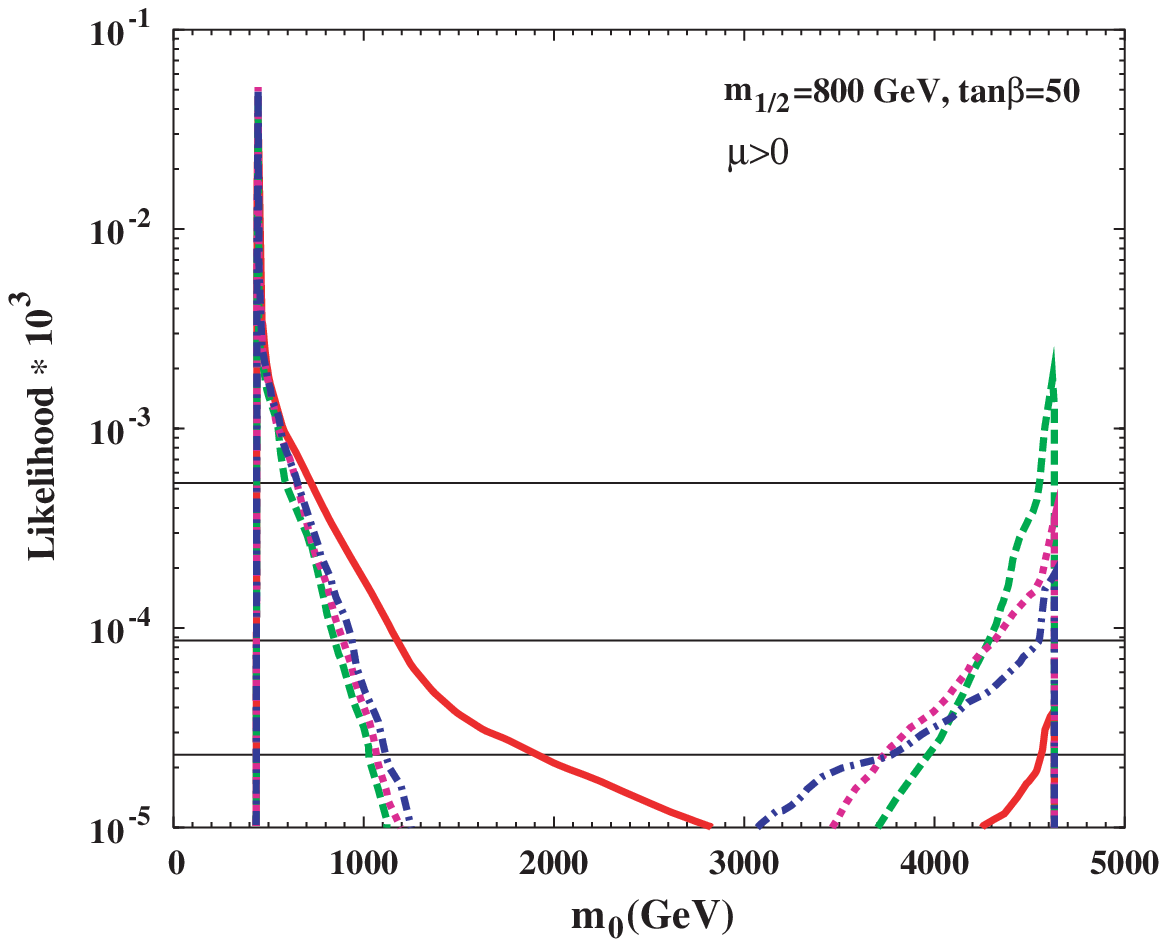,height=6cm}}
\mbox{\epsfig{file=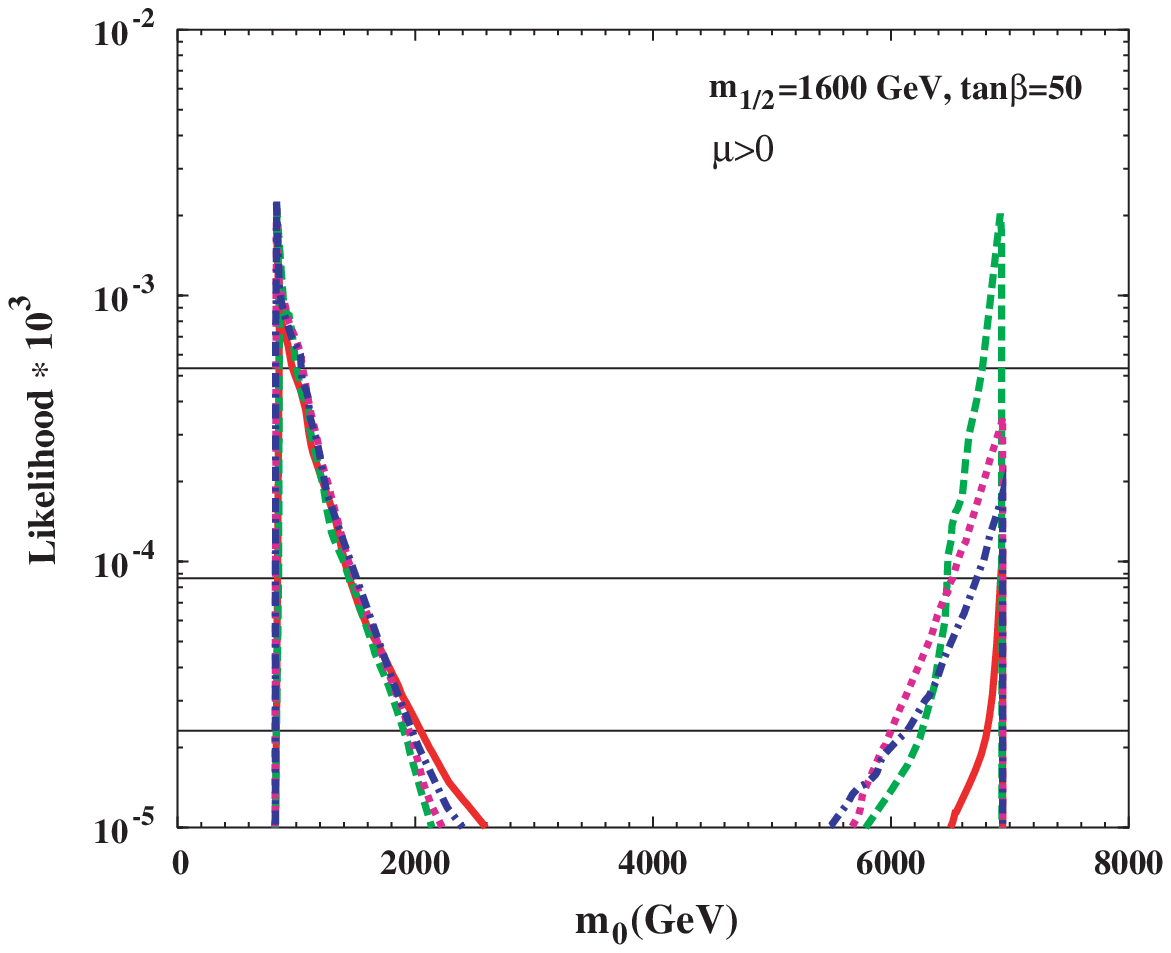,height=6cm}}
\end{center} 
\caption{\it
As in Fig.~\protect\ref{fig:WMAPFP} for $\tan \beta = 50, A_0 = 0, \mu > 0$ 
and $m_{1/2} =  800, 1600$~GeV in the left and right panels.
The $g_\mu-2$ constraint is included (excluded) in the
bottom (top) panels. }
\label{fig:WMAP50}
\end{figure}

Fig.~\ref{fig:WMAP50} displays the likelihood function along cuts in the
$(m_{1/2}, m_0)$ plane for $\tan \beta = 50, A_0 = 0, \mu > 0$ and
$m_{1/2} = 800$~GeV (left panels) or $1600$~GeV (right panels). The $g_\mu
- 2$ contribution to the likelihood is included in the bottom panels, but not
in the top panels. The line styles are the same as in
Fig.~\ref{fig:WMAPFP}, and we note that the coannihilation and focus-point
regions even link up somewhat below the 95\% CL in the case of $m_0 =
800$~GeV, if the present error in $m_t$ is assumed, but only if the
$g_\mu - 2$ contribution to the likelihood is discarded. In this case, we can
not resolve the difference between the coannihilation and funnel peaks.

\section{Likelihood Contours in the $(m_{1/2}, m_0)$ Planes}
\label{sec:contours}

Using the fully normalized likelihood function ${\cal L}_{tot}$ obtained
by combining both signs of $\mu$ for each value of $\tan \beta$, we now
determine the regions in the $(m_{1/2}, m_0)$ planes which correspond to
specific CLs. For a given CL, $x$, an iso-likelihood contour is determined
such that the integrated volume of ${\cal L}_{tot}$ within that contour is
equal to $x$, when the total volume is normalized to unity. The values of
the likelihood corresponding to the displayed contours are tabulated in
Table \ref{table:cl} (with $g_{\mu} - 2$) and Table \ref{table:cl2} (without $g_{\mu} - 2$).

\begin{figure}
\begin{center}
\mbox{\epsfig{file=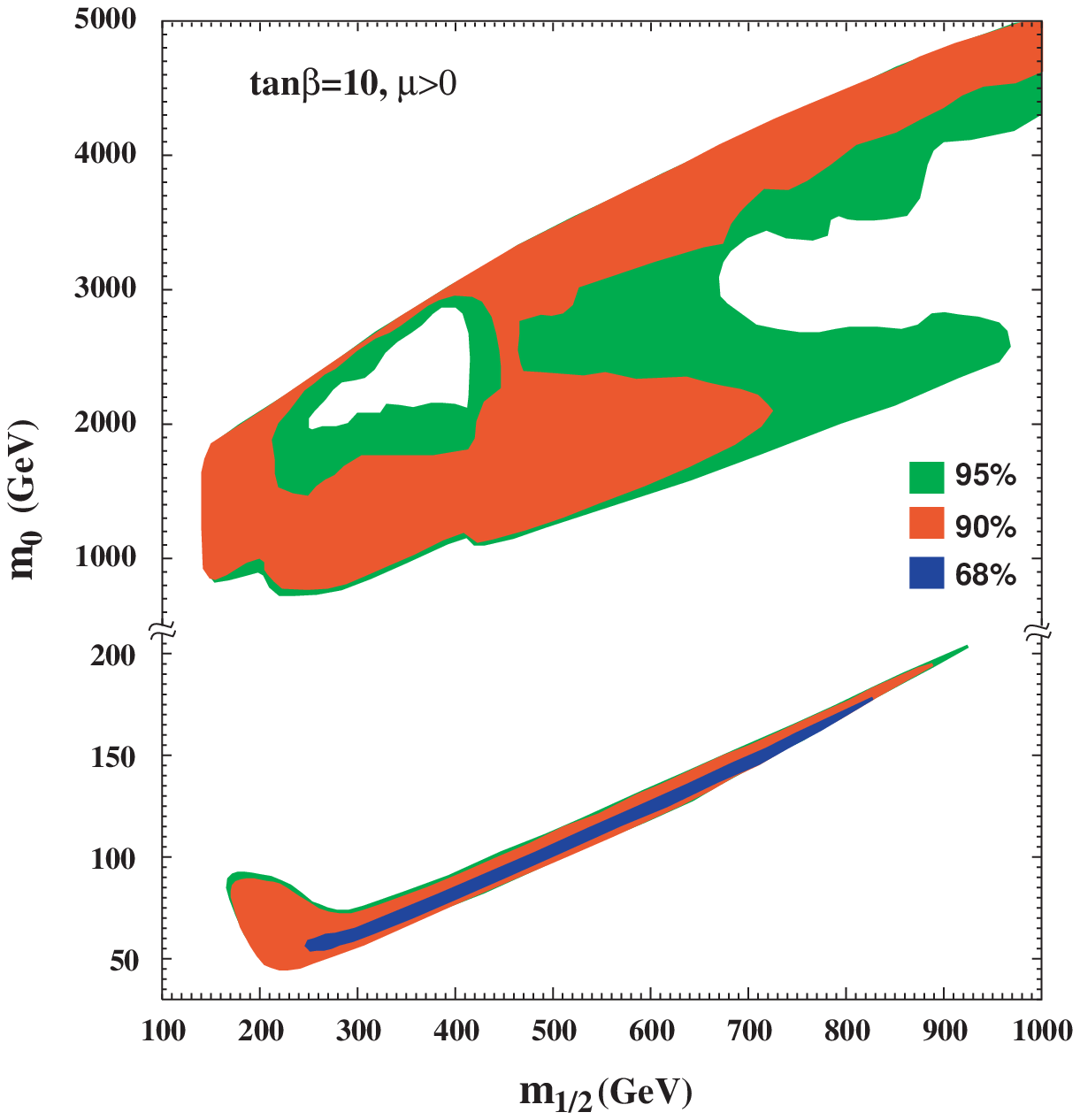,height=8cm}}
\mbox{\epsfig{file=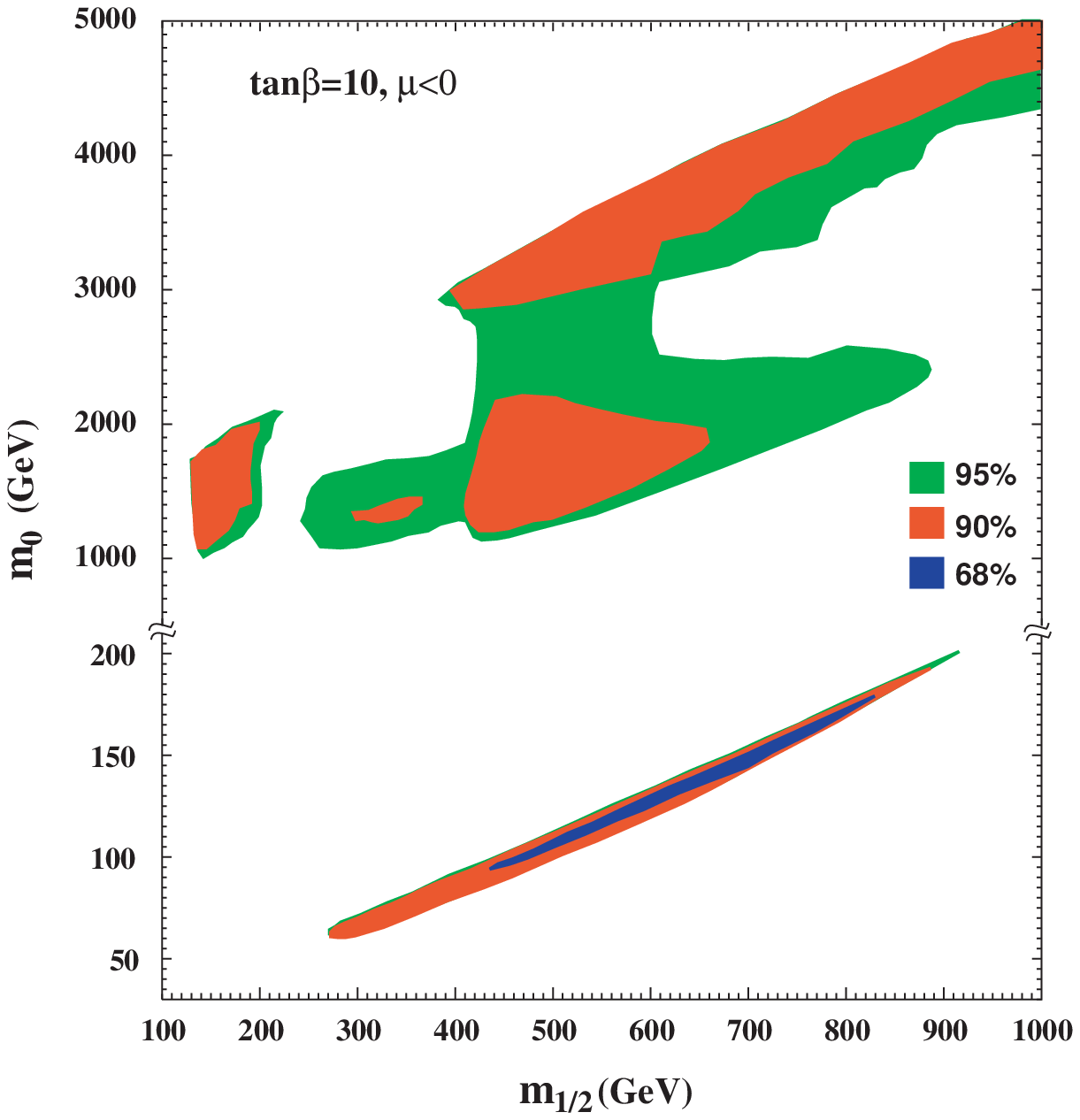,height=8cm}}
\end{center}   
\caption{\it
Contours of the likelihood at the 68\%, 90\% and 95\% levels for $\tan 
\beta = 10$, $A_0 = 0$ and $\mu > 0$ (left panel) or $\mu < 0$ (right 
panel), calculated 
using information of $m_h$, $b \to s \gamma$ and $\Omega_{CDM} h^2$ and 
the current uncertainties in $m_t$ and $m_b$, but without using any 
information about $g_\mu - 2$.
}
\label{fig:contours}   
\end{figure}

Fig.~\ref{fig:contours} extends the previous analysis to the entire
$(m_{1/2}, m_0)$ plane for $\tan \beta = 10$ and $A_0 = 0$, including both
signs of $\mu$. The darkest (blue), intermediate (red) and lightest
(green) shaded regions are, respectively, those where the likelihood is
above 68\%, above 90\%, and  above 95\%. 
Overall, the likelihood for $\mu < 0$ is less than that for 
$\mu > 0$, even without including any information about $g_\mu - 2$ 
due to the Higgs and $b \to s \gamma$ constraints. 
Only the bulk and coannihilation-tail regions appear above the 68\% level, 
but the focus-point region appears above the 90\% level, and so cannot be 
excluded. 

The
highly non-Gaussian behaviour of the likelihood shown in
Fig.~\ref{fig:contours} can be understood when comparing this figure to
Fig.~\ref{fig:WMAPFP}(a,b). At fixed $m_{1/2}$ and for a given CL,
portions of the likelihood function above the horizontal lines in
\ref{fig:WMAPFP}(a,b) correspond to shaded regions in
Fig.~\ref{fig:contours}. The broad low-lying bump or plateau in the
likelihood function at intermediate values of $m_0$ is now reflected in
the extended features seen in Fig.~\ref{fig:contours}. The extent of this
plateau is somewhat diminished for $\mu < 0$.

\begin{figure}
\begin{center}
\mbox{\epsfig{file=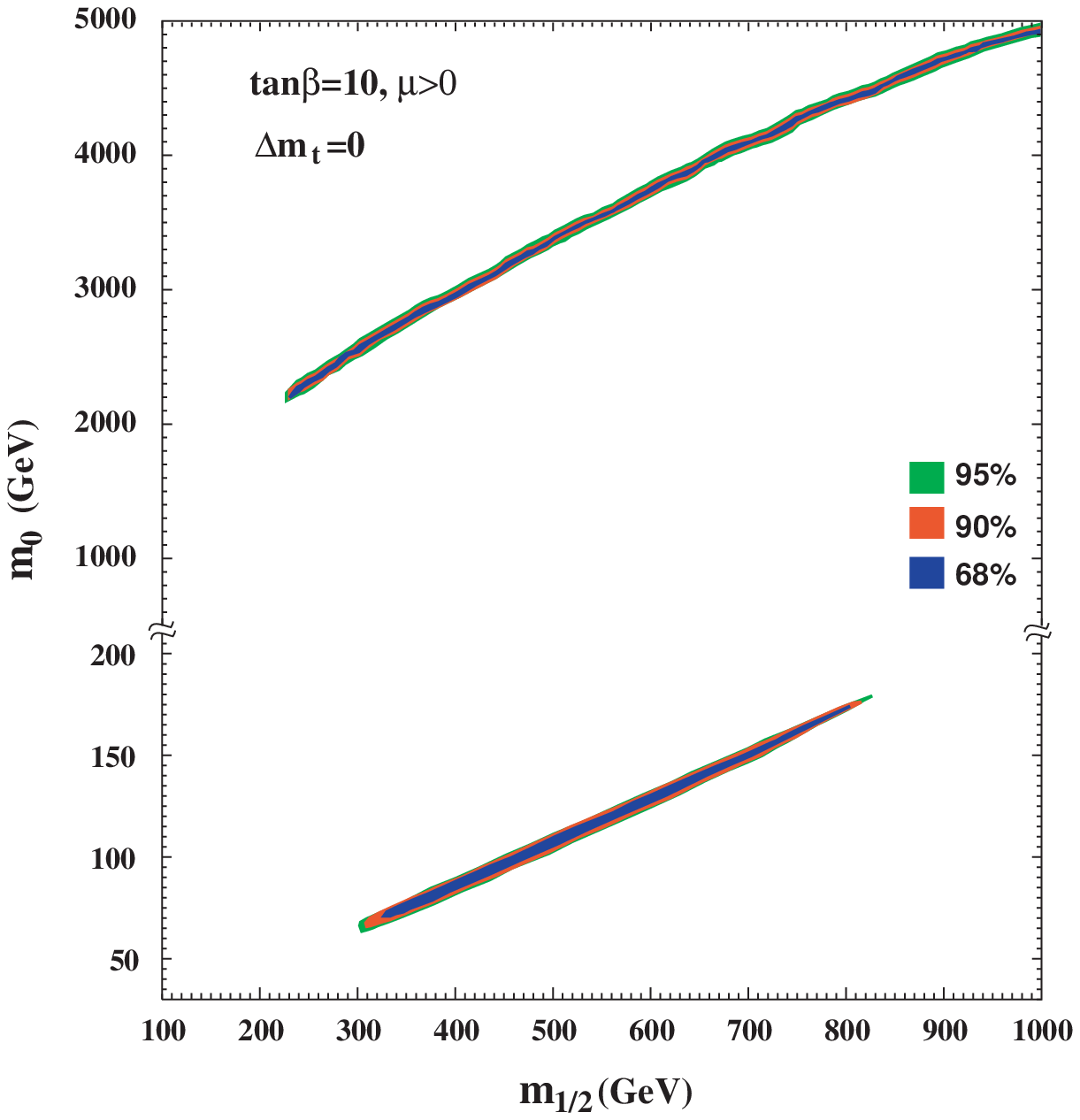,height=8cm}}
\mbox{\epsfig{file=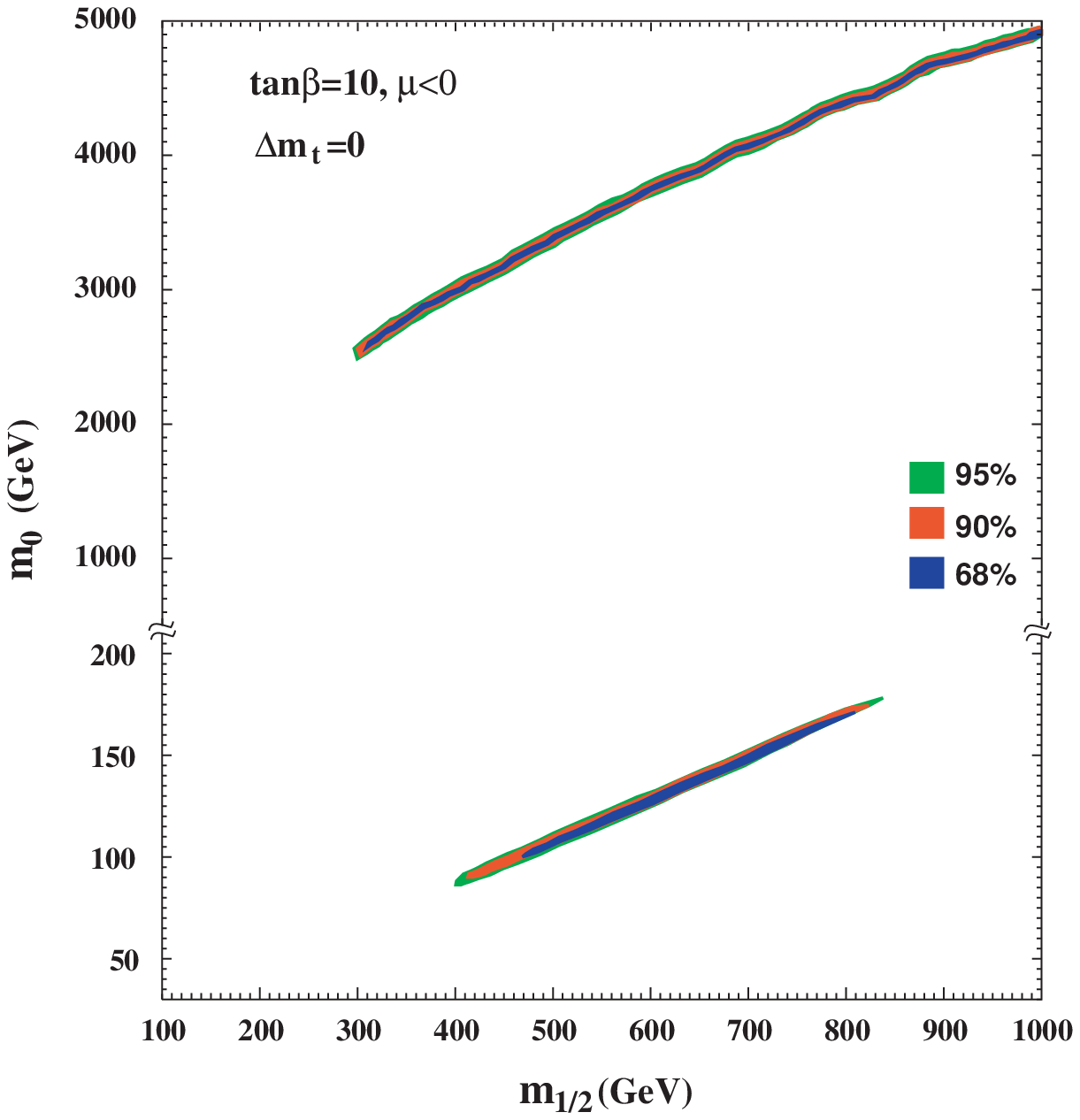,height=8cm}}
\end{center}   
\caption{\it
As in Fig.~\protect\ref{fig:contours} but assuming zero uncertainty in 
$m_t$.
}
\label{fig:contourswithoutmt}   
\end{figure}

The bulk region is more apparent in the left panel of
Fig.~\ref{fig:contours} for $\mu > 0$ than it would be if the experimental
error in $m_t$ and the theoretical error in $m_h$ were neglected. 
Fig.~\ref{fig:contourswithoutmt} complements the previous figures by 
showing the likelihood functions as they would appear if there were no 
uncertainty in $m_t$, keeping the other inputs the same and using no 
information about $g_\mu - 2$. We see that, in this case, both the 
coannihilation and focus-point strips rise above the 68\% CL.

\begin{figure}
\begin{center}
\mbox{\epsfig{file=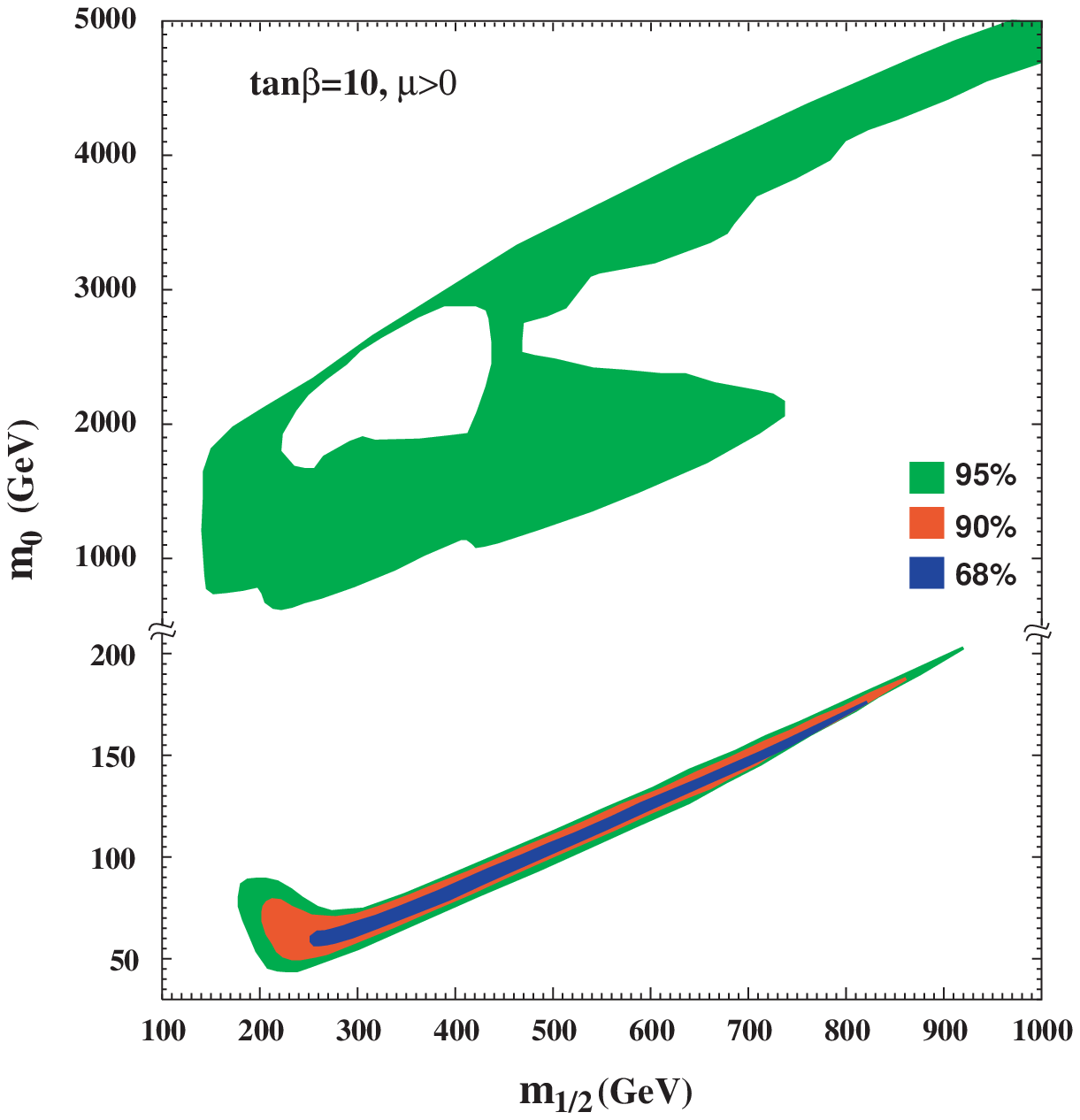,height=8cm}}
\mbox{\epsfig{file=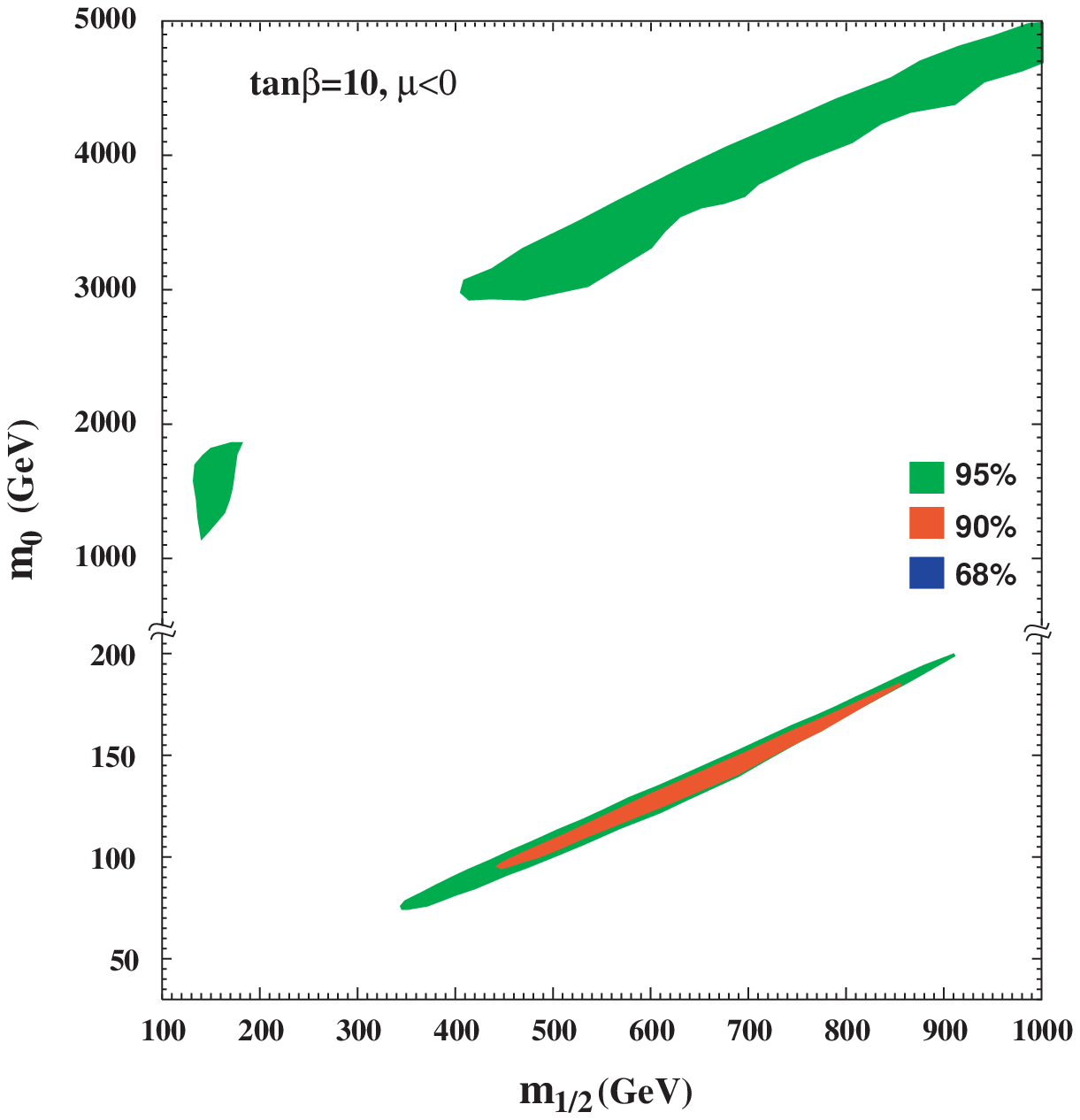,height=8cm}}
\end{center}   
\caption{\it
Likelihood contours as in Fig.~\ref{fig:contours}, but also including the 
information obtained by 
comparing the experimental measurement of $g_\mu - 2$ with the Standard 
Model estimate based on $e^+ e^-$ data.} 
\label{fig:contourswithmu}   
\end{figure}

Fig.~\ref{fig:contourswithmu} is also for $\tan \beta = 10$ and $A_0 = 0$,
including both signs of $\mu$. This time, we include also the $g_\mu - 2$
likelihood, calculated on the basis of the $e^+ e^-$ annihilation estimate
of the Standard Model contribution. This figure represents an extension of
Fig.~\ref{fig:WMAPFP}(c,d). In this case, very low values of $m_{1/2}$ and 
$m_0$ are disfavoured when
$\mu > 0$. Furthermore, for $\mu < 0$ the likelihood is suppressed and
no part of the coannihilation tail is above the 68\% CL. In addition,
none of the focus point region lies above the 90\% CL for either positive
or negative $\mu$.  However, neither
of these can be excluded completely, since there are $\mu < 0$ zones
within the 90\% likelihood contour, and focus-point zones within the 95\%
likelihood contour.

\begin{figure}
\begin{center}
\mbox{\epsfig{file=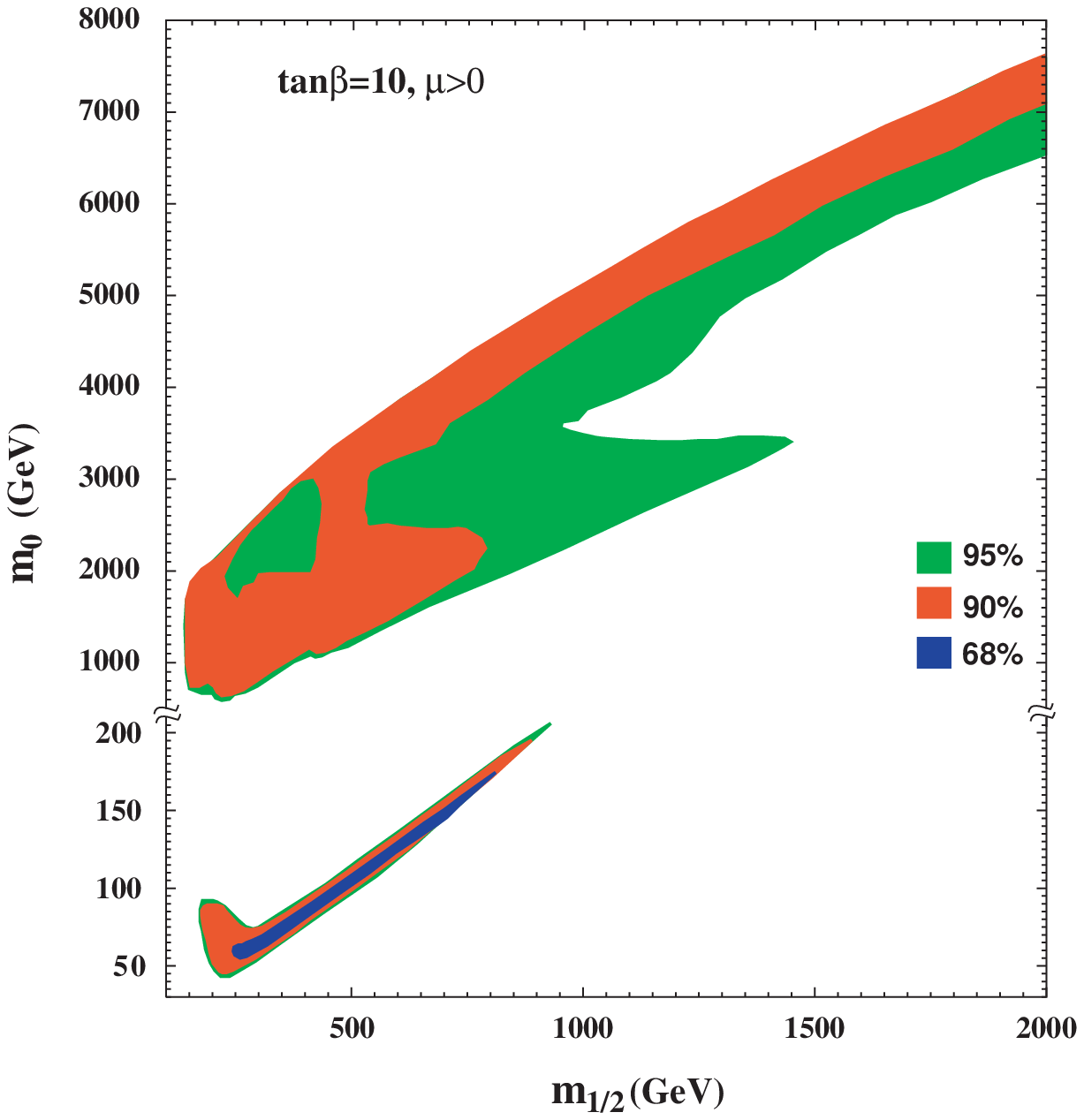,height=8cm}}
\mbox{\epsfig{file=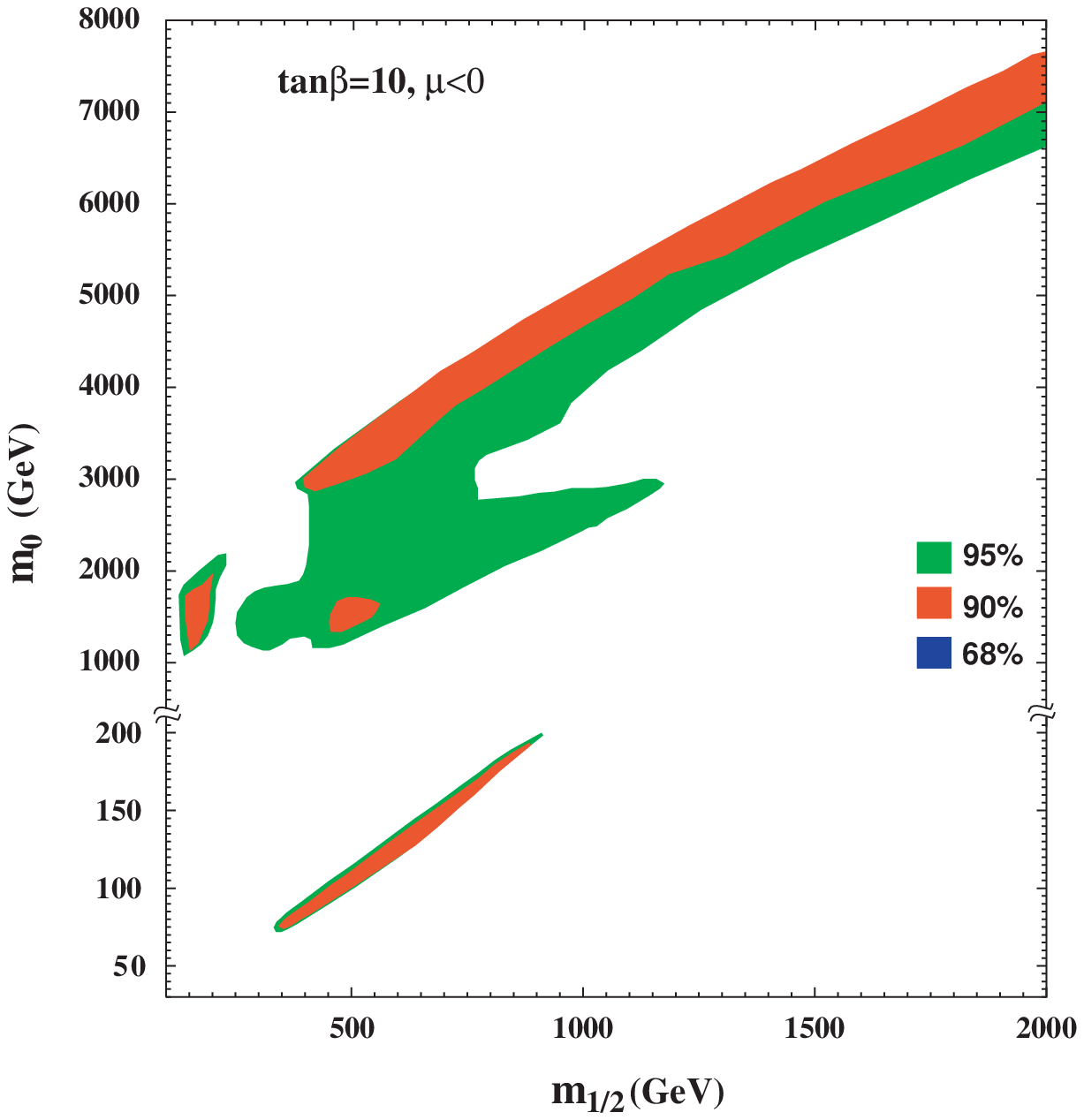,height=8cm}}
\end{center}   
\caption{\it
Likelihood contours as in Fig.~\ref{fig:contourswithmu}, but extending the
range for $m_{1/2}$ up to 2~TeV.} 
\label{fig:highm12}   
\end{figure}

It is important to note that the results presented thus far are somewhat
dependent on the range chosen for $m_{1/2}$, which has so far been
restricted for $\tan \beta = 10$ to $\leq 1$ TeV. In
Fig.~\ref{fig:highm12}, we show the the $\tan \beta = 10$ plane for $\mu >
0$ and $\mu < 0$ allowing $m_{1/2}$ up to 2 TeV, including the $g_\mu - 2$
constraint. Comparing this figure with Fig.~\ref{fig:contourswithmu}, we
see that a considerable portion of the focus-point region is now above the
90\% CL due to the enhanced volume contribution at large $m_{1/2}$.

\begin{figure}
\begin{center}
\mbox{\epsfig{file=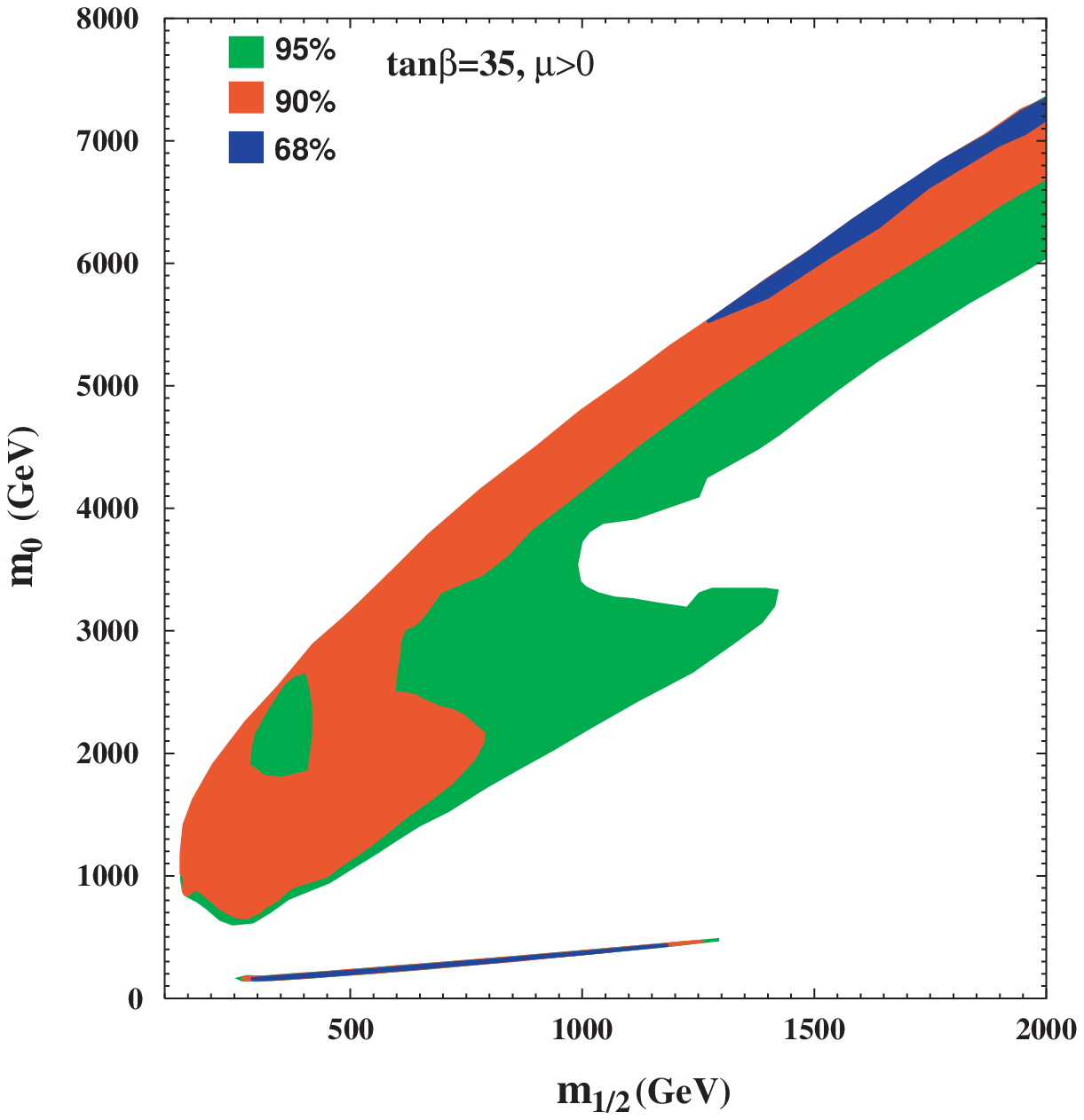,height=8cm}}
\mbox{\epsfig{file=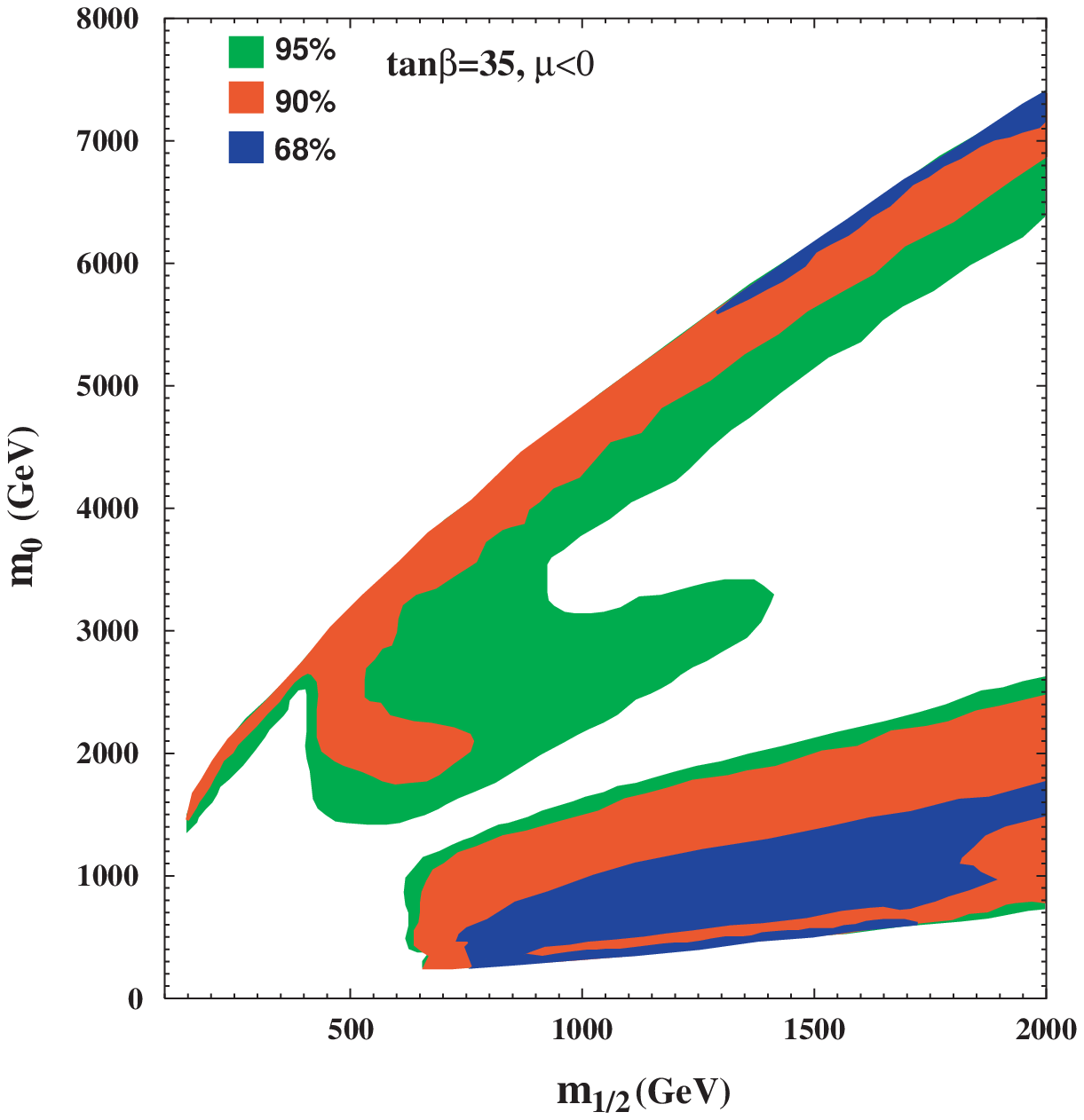,height=8cm}}
\end{center}   
\caption{\it
Likelihood contours as in Fig.~\ref{fig:contours}, but for $\tan 
\beta = 35$, $A_0 = 0$ and $\mu> 0$ ($\mu < 0$) in the left (right)
panel, calculated 
using information of $m_h$, $b \to s \gamma$ and $\Omega_{CDM} h^2$ and 
the current uncertainties in $m_t$ and $m_b$, without 
the indicative information from $g_\mu - 2$.
}
\label{fig:contours35}   
\end{figure}

\begin{figure}
\begin{center}
\mbox{\epsfig{file=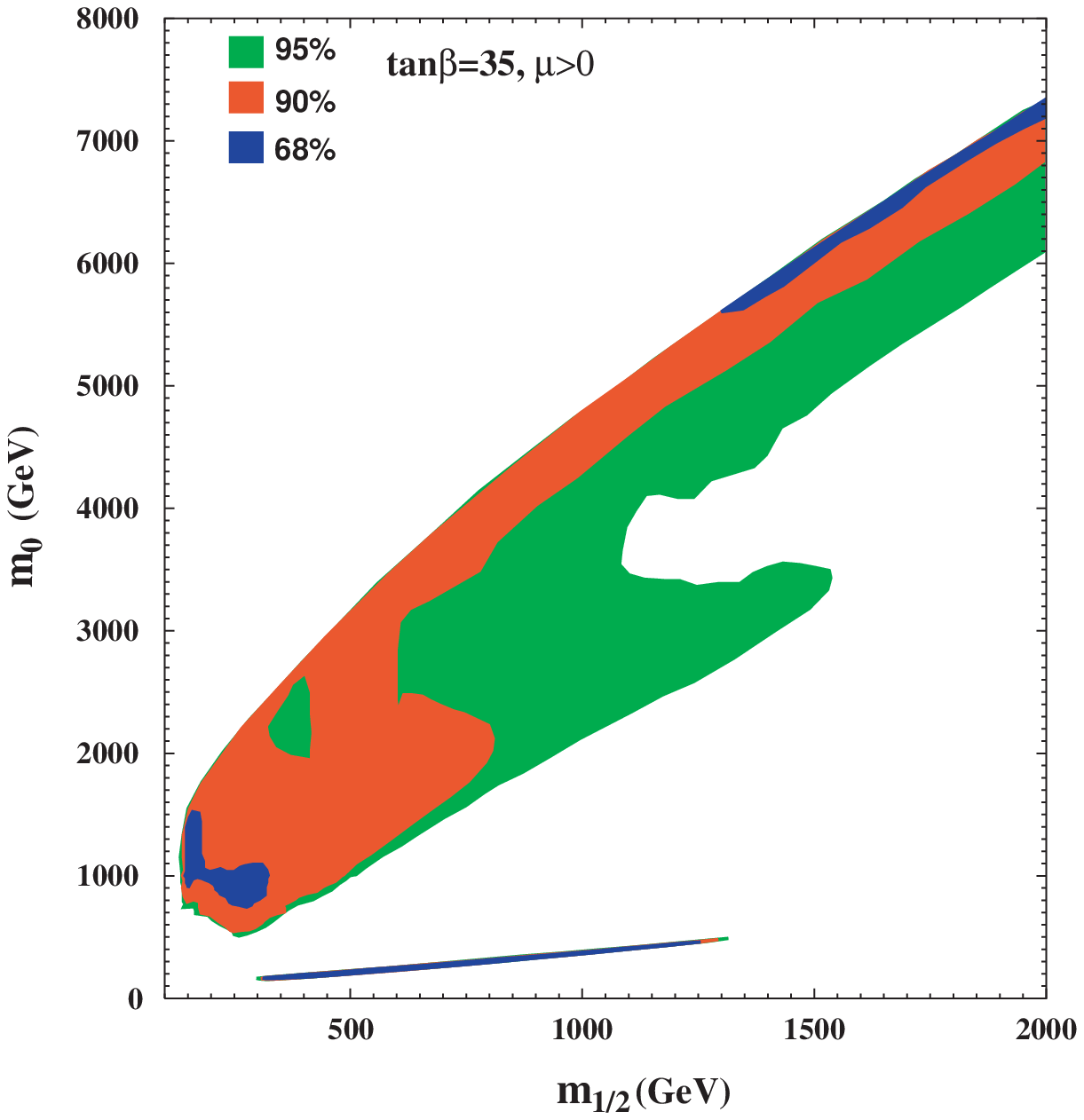,height=8cm}}
\mbox{\epsfig{file=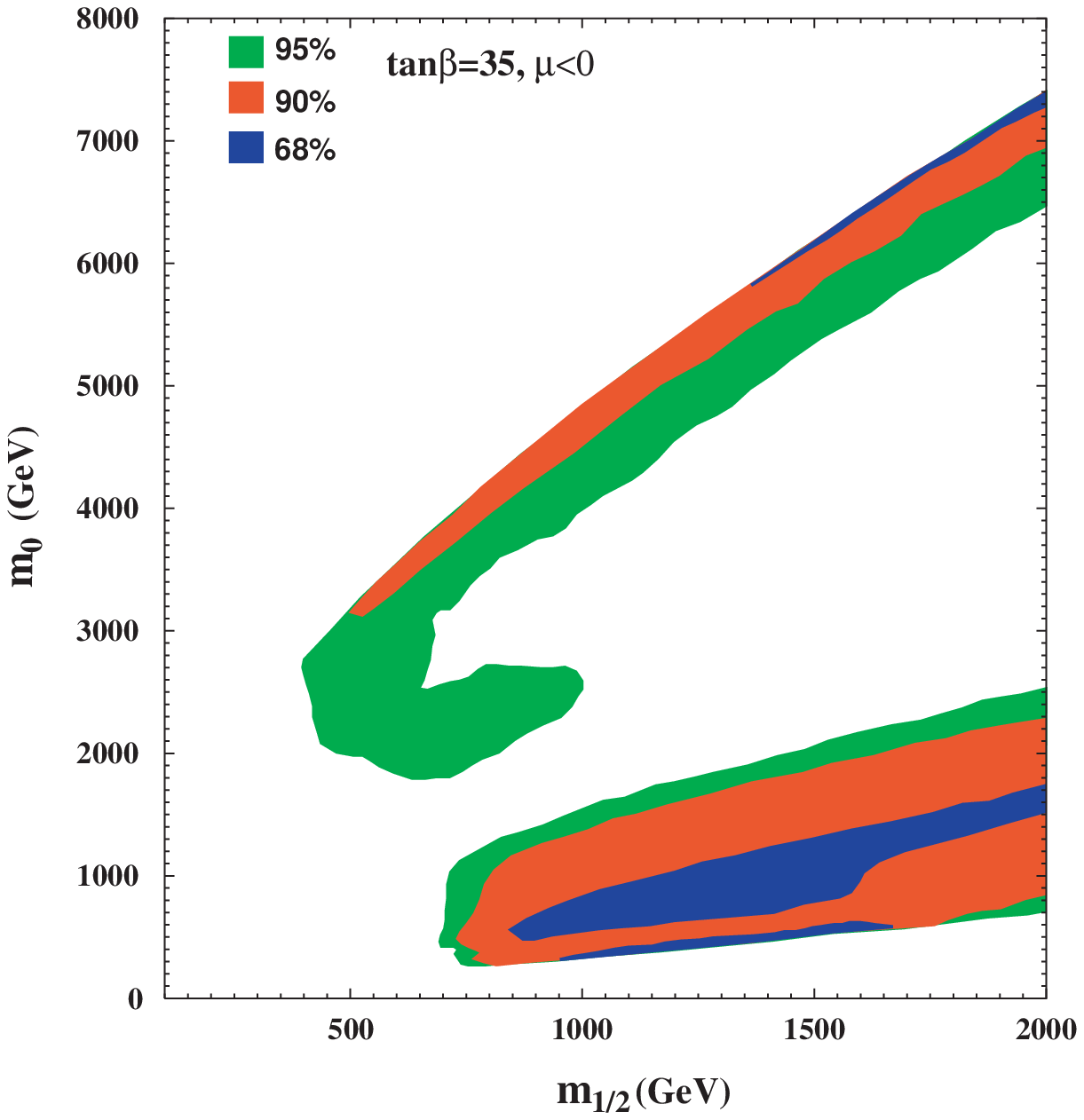,height=8cm}}
\end{center}   
\caption{\it
Likelihood contours as in Fig.~\ref{fig:contours35}, including
the indicative information from $g_\mu - 2$.
}
\label{fig:contours35withmu}   
\end{figure}

Fig.~\ref{fig:contours35} is for $\tan \beta = 35$, $A_0 = 0$  for both $\mu > 0 $ and 
$\mu < 0$. 
Fig.~\ref{fig:contours35withmu} includes the $g_\mu - 2$ likelihood, calculated on the
basis of the $e^+ e^-$ annihilation estimate of the Standard Model
contribution, which is not included in the previous figure.  In this case,
regions at small $m_{1/2}$ and $m_0$ are disfavoured by the $b \to s
\gamma$ constraint, as seen in both figures with $\mu < 0$. At larger $m_{1/2}$, the
coannihilation region is broadened by a merger with the rapid-annihilation
funnel that appears for large $\tan \beta$. The optional $g_\mu - 2$
constraint would prefer $\mu > 0$, and in the $\mu < 0$ half-plane it
favours larger $m_{1/2}$ and $m_0$, as seen when the left and right 
panels are compared.

\begin{figure}
\begin{center}
\mbox{\epsfig{file=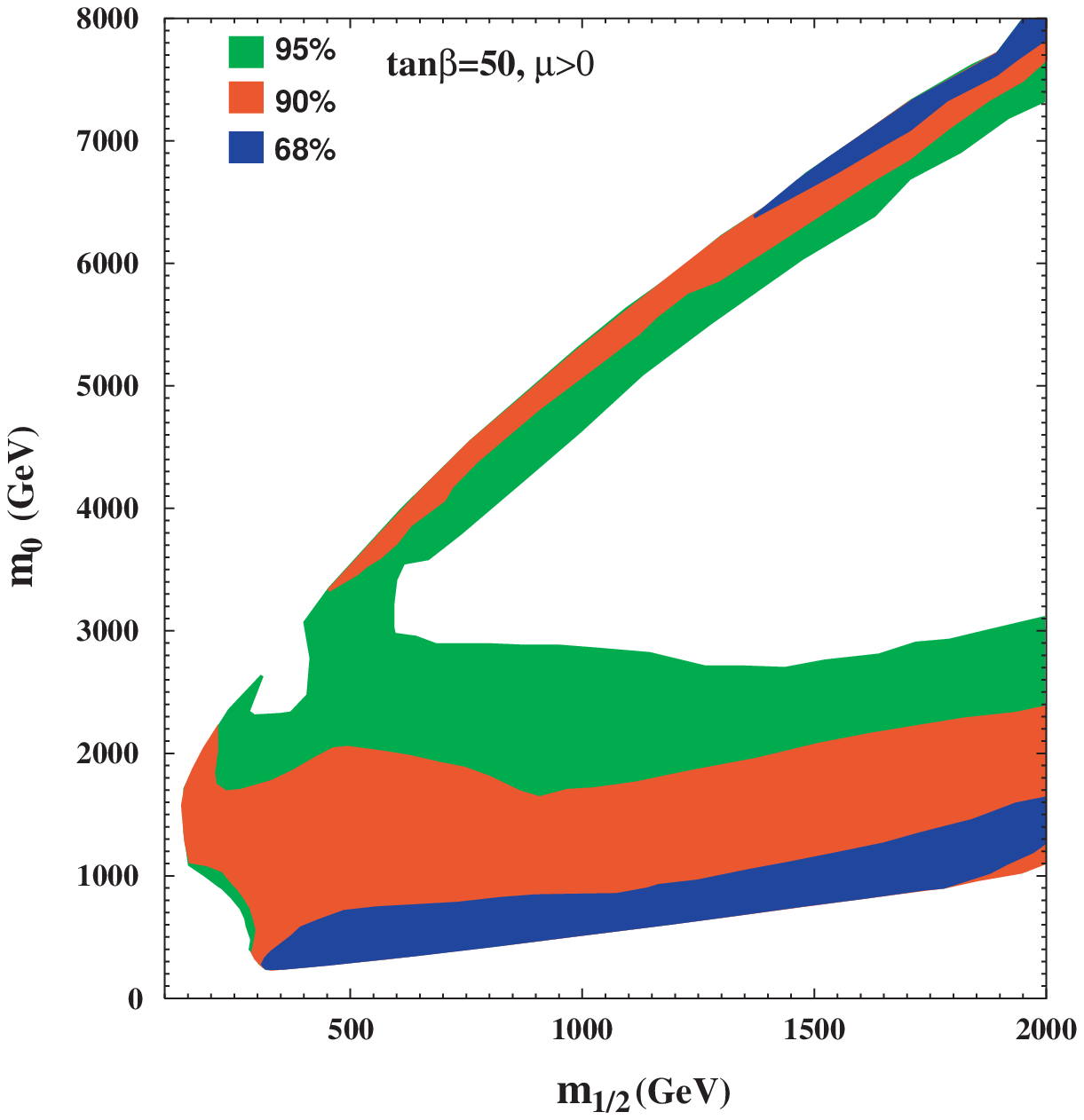,height=8cm}}
\mbox{\epsfig{file=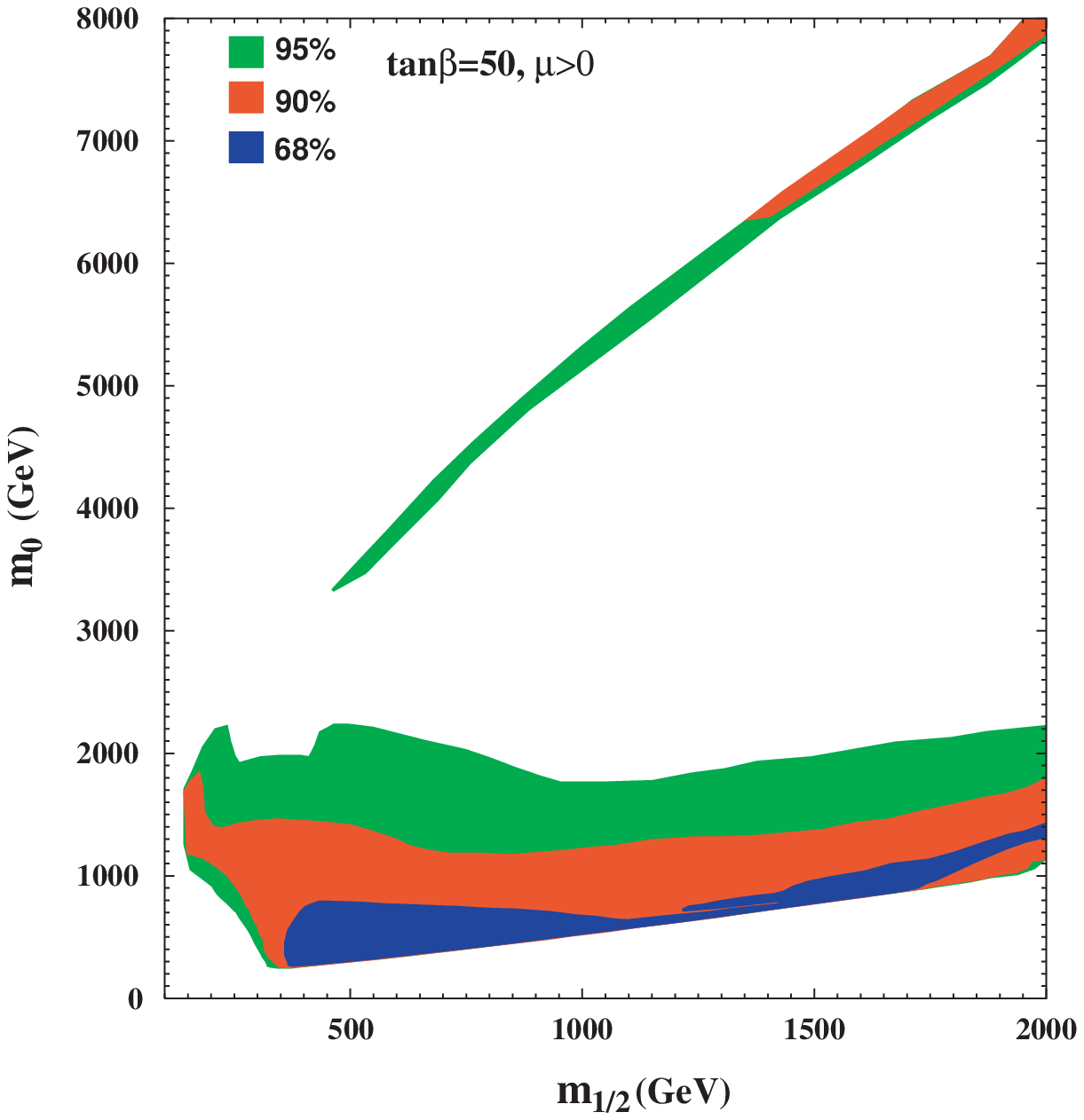,height=8cm}}
\end{center}   
\caption{\it
Likelihood contours as in Fig.~\ref{fig:contours35}, but for $\tan 
\beta = 50$, $A_0 = 0$ and $\mu > 0$, without (left panel) 
and with (right panel) the indicative information from $g_\mu - 2$.
}
\label{fig:contours50}   
\end{figure}

Fig.~\ref{fig:contours50} is for $\tan \beta = 50$, $A_0 = 0$ and $\mu >
0$. Again, the right panel includes the $g_\mu - 2$ likelihood, calculated
on the basis of the $e^+ e^-$ annihilation estimate of the Standard Model
contribution, which is not included in the left panel.  In this case, the
disfavouring of regions at small $m_{1/2}$ and $m_0$ by the $b \to s
\gamma$ constraint is less severe than in the case of $\tan \beta = 35$
and $\mu < 0$, but is still visible in both panels.  The coannihilation
region is again broadened by a merger with the rapid-annihilation funnel.
In the absence of the $g_\mu - 2$ constraint, both the coannihilation and
the focus-point regions feature strips allowed at the 68\% CL, and these
are linked by a bridge at the 95\% CL. However, when the optional $g_\mu -
2$ constraint is applied, this bridge disappears, the 90\% and 95\% CL
strips in the focus-point region becomes much thinner, and the 68\% strip
disappears in this region.

\section{Summary}
\label{sec:summary}

We have presented in this paper a new global likelihood analysis of the
CMSSM, incorporating the likelihoods contributed by $m_h$, $b \to s
\gamma$, $\Omega_{CDM} h^2$ and (optionally) $g_\mu - 2$. We have
discussed extensively the impacts of the current experimental
uncertainties in $m_t$ and $m_b$, which affect each of $m_h$, $b \to s
\gamma$ and $\Omega_{CDM} h^2$. In particular, the widths of the
coannihilation and focus-point strips are sensitive to the uncertainties
in $m_t$ and $m_b$, and a low-lying plateau in the likelihood is found
with the present uncertainty $\Delta m_t = 5$~GeV.

We recall that the absolute values of the likelihood integrals are not in
themselves meaningful, but their relative values do carry some
information.  Generally speaking, the global likelihood function reaches
higher values in the coannihilation region than in the focus-point region,
as can be seen by comparing the entries with and without parentheses in
Table~\ref{table:like}. This tendency would have been reversed if the
uncertainty in $m_t$ had been neglected, as seen in
Table~\ref{table:like_dmt0}, but the preference for the coannihilation
region is in any case not conclusive.

Table~\ref{table:like} also displays the integrated likelihood function
for different values of $\tan \beta$ and the sign of $\mu$, exhibiting a
weak general preference for $\mu > 0$ if the $g_\mu - 2$ information is
used. If this information is not used, $\mu < 0$ is preferred for $\tan
\beta = 35$, but $\mu > 0$ is still preferred for $\tan \beta = 10$. There
is no significant preference for any value between $\tan \beta = 10$ and
the upper limits $\gappeq 35$ and $\gappeq 50$ where electroweak symmetry
breaking ceases to be possible in the CMSSM, though we do find a weak
preference for $\tan \beta = 50$ and $\mu > 0$.

\begin{table}
\begin{center}
\caption{\it Integrals of the global likelihood function 
integrated over  the $(m_{1/2},m_0)$ planes for various
values of $\tan\beta$, in the coannihilation and funnel (focus-point) 
region.
We use the range $m_{1/2} \leq 2\, TeV$, except for the second row of 
$\tan\beta =10$ 
case, where the range $m_{1/2} \leq 1\, TeV$ is used.}
\vspace*{3mm}
\label{table:like}
\begin{tabular}{|c|c|c||c|c|}
\hline\hline
& \multicolumn{2}{c||}{incl. $g_\mu-2$} & \multicolumn{2}{c|}{without $g_\mu-2$}  \\ \cline{2-5}
\multicolumn{1}{|c|}{$\tan\beta$}  & $\mu>0$ & $\mu<0$ & $\mu>0$ & $\mu<0$  \\ \hline
10   &  41.7 (5.9)   & 2.1 (4.8)    &  2329 (1052)   & 1147 (984)      \\ 
     &  41.7 (2.9)   & 2.1 (1.7)    &  2329 (476)    & 1147 (387)      \\ \hline
35   &  33.9 (11.6)  & 25.9 (5.5)   &  1428 (1596)   &  8690 (1270)    \\ \hline
50   &  231.9 (6.84) &              &  13096 (1117)  &                 \\ \hline
\hline
\end{tabular}
\end{center}
\end{table}

\begin{table}
\begin{center}
\caption{\it As in Table~\ref{table:like}, but assuming zero uncertainty in $m_t$.}
\vspace*{3mm}
\label{table:like_dmt0}
\begin{tabular}{|c|c|c||c|c|}
\hline\hline
& \multicolumn{2}{c||}{incl. $g_\mu-2$} & \multicolumn{2}{c|}{without $g_\mu-2$}  \\ \cline{2-5}
\multicolumn{1}{|c|}{$\tan\beta$}  & $\mu>0$ & $\mu<0$ & $\mu>0$ & $\mu<0$  \\ \hline
10   &   44.9 (69.1)  &    2.6 (67.7)    &  2425 (12916)  &  1485 (13442)    \\ 
     &   44.9 (21.4)  &    2.6 (20.2)    &  2425 (3922) &  1485 (4144)  \\ \hline
35   &   33.5 (90.9)  &   26.4 (58.3)    &  1451 (15377) &  8837 (12589) \\ \hline
50   &   195.0 (60.4) &                  &  13877 (10188)&              \\ \hline
\hline
\end{tabular}
\end{center}
\end{table}

In the foreseeable future, the analysis in this paper could be refined
with the aid of improved measurements of $m_t$ at the Fermilab Tevatron
collider, by refined estimates of $m_b$, by better determinations of
$\Omega_{CDM} h^2$ and more experimental and theoretical insight into
$g_\mu - 2$, in particular. One could also consider supplementing our
phenomenological analysis with arguments based on naturalness or
fine-tuning, which would tend to disfavour larger values of $m_{1/2}$ and
$m_0$.  However, in the absence of such theoretical arguments, our
analysis shows that long strips in the coannihilation and focus-point
regions cannot be excluded on the basis of present data. The preparations
for searches for supersymmetry at future colliders should therefore not be
restricted to low values of $m_{1/2}$ and $m_0$.

\vskip 0.5in
\vbox{
\noindent{ {\bf Acknowledgments} } \\
\noindent 
We thank Martin Gr\"unewald and Peter Igo-Kemenes for help with the Higgs 
likelihood, and Geri Ganis for help with the $b \to s \gamma$ 
likelihood. The work of K.A.O., Y.S., and V.C.S. was supported in part
by DOE grant DE--FG02--94ER--40823.}


\begin{thebibliography}{99}

\bibitem{LEPHWG}
LEP Higgs Working Group for Higgs boson searches, OPAL Collaboration,
ALEPH Collaboration, DELPHI Collaboration and L3
Collaboration,
Phys.\ Lett.\ B {\bf 565} (2003) 61 [arXiv:hep-ex/0306033].
{\it Searches for the neutral Higgs bosons of the MSSM: Preliminary
combined results using LEP data collected at energies up to 209 GeV},
LHWG-NOTE-2001-04, ALEPH-2001-057, DELPHI-2001-114, L3-NOTE-2700,
OPAL-TN-699, arXiv:hep-ex/0107030; LHWG Note/2002-01,\\
{\tt 
http://lephiggs.web.cern.ch/LEPHIGGS/papers/July2002{\_}SM/index.html}.

\bibitem{bsgex}
S.~Chen {\it et al.}  [CLEO Collaboration],
Phys.\ Rev.\ Lett.\  {\bf 87} (2001) 251807
[arXiv:hep-ex/0108032];
BELLE Collaboration, BELLE-CONF-0135.
See also
K.~Abe {\it et al.}  [Belle Collaboration],
Phys.\ Lett.\ B {\bf 511} (2001) 151 [arXiv:hep-ex/0103042];
B.~Aubert {\it et al.}  [BaBar Collaboration],
arXiv:hep-ex/0207076.

\bibitem{BNL}
G.~W.~Bennett {\it et al.}  [Muon g-2 Collaboration],
Phys.\ Rev.\ Lett.\  {\bf 89} (2002) 101804
[Erratum-ibid.\  {\bf 89} (2002) 129903]
[arXiv:hep-ex/0208001].

\bibitem{newDavier}
M.~Davier, S.~Eidelman, A.~Hocker and Z.~Zhang,
arXiv:hep-ph/0308213.

\bibitem{ganis}
J.~R.~Ellis, T.~Falk, G.~Ganis, K.~A.~Olive and M.~Srednicki,
Phys.\ Lett.\ B {\bf 510} (2001) 236
[arXiv:hep-ph/0102098].



\bibitem{LS}
A.~B.~Lahanas and V.~C.~Spanos,
Eur.\ Phys.\ J.\ C {\bf 23} (2002) 185
[arXiv:hep-ph/0106345].

\bibitem{eos2}
J.~R.~Ellis, K.~A.~Olive and Y.~Santoso,
New J.\ Phys.\  {\bf 4} (2002) 32
[arXiv:hep-ph/0202110].



\bibitem{cmssm} Some additional recent papers include:
V.~D.~Barger and C.~Kao,
Phys.\ Lett.\ B {\bf 518} (2001) 117
[arXiv:hep-ph/0106189];
L.~Roszkowski, R.~Ruiz de Austri and T.~Nihei,
JHEP {\bf 0108} (2001) 024
[arXiv:hep-ph/0106334];
A.~Djouadi, M.~Drees and J.~L.~Kneur,
JHEP {\bf 0108} (2001) 055
[arXiv:hep-ph/0107316];
U.~Chattopadhyay, A.~Corsetti and P.~Nath,
Phys.\ Rev.\ D {\bf 66} (2002) 035003
[arXiv:hep-ph/0201001];
H.~Baer, C.~Balazs, A.~Belyaev, J.~K.~Mizukoshi, X.~Tata and Y.~Wang,
JHEP {\bf 0207} (2002) 050
[arXiv:hep-ph/0205325];
R.~Arnowitt and B.~Dutta,
arXiv:hep-ph/0211417;
J.~R.~Ellis, K.~A.~Olive, Y.~Santoso and V.~C.~Spanos,
Phys.\ Lett.\ B {\bf 573} (2003) 163
[arXiv:hep-ph/0308075].





\bibitem{wmap}
C.~L.~Bennett {\it et al.},
Astrophys.\ J.\ Suppl.\  {\bf 148} (2003) 1
[arXiv:astro-ph/0302207].



\bibitem{reion}
D.~N.~Spergel {\it et al.},
Astrophys.\ J.\ Suppl.\  {\bf 148} (2003) 175
[arXiv:astro-ph/0302209].


\bibitem{EOSS}
J.~R.~Ellis, K.~A.~Olive, Y.~Santoso and V.~C.~Spanos,
Phys.\ Lett.\ B {\bf 565} (2003) 176
[arXiv:hep-ph/0303043].


\bibitem{LN}
A.~B.~Lahanas and D.~V.~Nanopoulos,
Phys.\ Lett.\ B {\bf 568} (2003) 55
[arXiv:hep-ph/0303130].


\bibitem{Baer}
H.~Baer and C.~Balazs,
JCAP {\bf 0305} (2003) 006
[arXiv:hep-ph/0303114].



\bibitem{Nath}
U.~Chattopadhyay, A.~Corsetti and P.~Nath,
Phys.\ Rev.\ D {\bf 68} (2003) 035005
[arXiv:hep-ph/0303201].


\bibitem{Arnowitt}
R.~Arnowitt, B.~Dutta and B.~Hu,
arXiv:hep-ph/0310103.


\bibitem{DeBoer}
W.~de Boer, M.~Huber, C.~Sander and D.~I.~Kazakov,
arXiv:hep-ph/0106311.

\bibitem{benchmark}
M.~Battaglia {\it et al.}, 
Eur. Phys. J. C {\bf 22} (2001) 535
[arXiv:hep-ph/0106204].



\bibitem{FeynHiggs}
S.~Heinemeyer, W.~Hollik and G.~Weiglein,
Comput.\ Phys.\ Commun.\  {\bf 124} (2000) 76
[arXiv:hep-ph/9812320];
S.~Heinemeyer, W.~Hollik and G.~Weiglein,
Eur.\ Phys.\ J.\ C {\bf 9} (1999) 343
[arXiv:hep-ph/9812472].


\bibitem{martin}
S.~P.~Martin,
Phys.\ Rev.\ D {\bf 67} (2003) 095012
[arXiv:hep-ph/0211366] and talk at SUSY03, Tucson, Arizona (2003).


\bibitem{gam}
C. Degrassi, P. Gambino and G.~F. Giudice,
JHEP {\bf 0012} (2000) 009 [arXiv:hep-ph/0009337],
as implemented by P. Gambino and G. Ganis.

\bibitem{bsgth}
M.~Carena, D.~Garcia, U.~Nierste and C.~E.~Wagner,
Phys. Lett. B {\bf 499} (2001) 141 
[arXiv:hep-ph/0010003]; 
D.~A.~Demir and K.~A.~Olive,
Phys.\ Rev.\ D {\bf 65} (2002) 034007
[arXiv:hep-ph/0107329];
T.~Hurth,
arXiv:hep-ph/0106050.



\bibitem{ftuning}
J.~R.~Ellis and K.~A.~Olive,
Phys.\ Lett.\ B {\bf 514} (2001) 114
[arXiv:hep-ph/0105004].


\bibitem{focus}
J.~L.~Feng, K.~T.~Matchev and T.~Moroi,
Phys.\ Rev.\ Lett.\  {\bf 84} (2000) 2322
[arXiv:hep-ph/9908309];
J.~L.~Feng, K.~T.~Matchev and T.~Moroi,
Phys.\ Rev.\ D {\bf 61} (2000) 075005
[arXiv:hep-ph/9909334];
J.~L.~Feng, K.~T.~Matchev and F.~Wilczek,
Phys.\ Lett.\ B {\bf 482} (2000) 388
[arXiv:hep-ph/0004043].

\bibitem{hbrsb}
K.~L.~Chan, U.~Chattopadhyay and P.~Nath,
Phys.\ Rev.\ D {\bf 58} (1998) 096004
[arXiv:hep-ph/9710473].

\bibitem{rs}
A.~Romanino and A.~Strumia,
Phys.\ Lett.\ B {\bf 487} (2000) 165
[arXiv:hep-ph/9912301].




\bibitem{LNM}
A.~B.~Lahanas, N.~E.~Mavromatos and D.~V.~Nanopoulos,
arXiv:hep-ph/0308251.


\bibitem{coann}
J.~R.~Ellis, T.~Falk and K.~A.~Olive,
Phys.\ Lett.\ B {\bf 444} (1998) 367
[arXiv:hep-ph/9810360];
J.~R.~Ellis, T.~Falk, K.~A.~Olive and M.~Srednicki,
Astropart.\ Phys.\  {\bf 13} (2000) 181
[Erratum-ibid.\  {\bf 15} (2001) 413]
[arXiv:hep-ph/9905481].


\end{thebibliography}
\end{document}